\documentclass[pdflatex,sn-mathphys-num]{sn-jnl}

\usepackage{graphicx}%
\usepackage{multirow}%
\usepackage{amsmath,amssymb,amsfonts}%
\usepackage{amsthm}%
\usepackage{mathrsfs}%
\usepackage[title]{appendix}%
\usepackage{xcolor}%
\usepackage{textcomp}%
\usepackage{manyfoot}%
\usepackage{booktabs}%
\usepackage{algorithm}%
\usepackage{algorithmicx}%
\usepackage{algpseudocode}%
\usepackage{listings}%
\usepackage{subcaption}

\theoremstyle{thmstyleone}%
\theoremstyle{thmstyletwo}%

\theoremstyle{thmstylethree}%

\begin{document}
	
	\title[Article Title]{Comprehensively stratifying MCIs into distinct risk subtypes based on brain imaging genetics fusion learning}
	
	
	\author[1]{\fnm{Muheng} \sur{Shang}}\email{shangmuheng@mail.nwpu.edu.cn}
	
	\author[1]{\fnm{Jin} \sur{Zhang}}\email{jinzhang@mail.nwpu.edu.cn}
	
	\author[1]{\fnm{Junwei} \sur{Han}}\email{jhan@nwpu.edu.cn}
	
	\author[1*]{\fnm{Lei} \sur{Du}}\email{dulei@nwpu.edu.cn}
	
	\affil*[1]{\orgdiv{School of Automation}, \orgname{Northwestern Polytechnical University}, \orgaddress{\street{127 Youyi Road}, \city{Xi'an}, \postcode{710072}, \state{Shaanxi}, \country{China}}}
	
%
	
	
\abstract{Mild cognitive impairment (MCI) is the prodromal stage of Alzheimer's disease (AD) and thus enrolling MCI subjects to undergo clinical trials is worthwhile. However, MCI groups usually show significant diversity and heterogeneity in the pathology and symptom, which pose great challenge to accurately select appropriate subjects. This study aimed to stratify MCI subjects into distinct subgroups with substantial difference in the risk of transitioning to AD by fusing multimodal brain imaging genetic data. The integrated imaging genetics method comprised three modules, i.e., the whole-genome-oriented risk genetic information extraction module (RGE), the genetic-to-brain mapping module (RG2PG), and the genetic-guided pseudo-brain fusion module (CMPF). We used data from AD Neuroimaging Initiative (ADNI) and identified two MCI subtypes, called low-risk MCI (lsMCI) and high-risk MCI (hsMCI). We also validated that the two subgroups showed distinct patterns of in terms of multiple biomarkers including genetics, demographics, fluid biomarkers, brain imaging features, clinical symptoms and cognitive functioning at baseline, as well as their longitudinal developmental trajectories. Furthermore, we also identified potential biomarkers that may implicate the risk of MCIs, providing critical insights for patient stratification at early stage.}

	
	
	
	\maketitle
	
\section{Introduction}\label{sec1}
AD is a severe type of dementia and cannot be cured once diagnosed. This suggests that the early intervention is by now recognized as the best way to slow AD development, which can prolong the disease progression and lower the cost of care \cite{knopman2021alzheimer}. Mild cognitive impairment (MCI) is generally accepted as the prodromal stage of AD, and many studies, with aim to intervene earlier, enrolled patients with MCI to undergo clinical trials \cite{portet2006mild}. According to National Institute on Aging and Alzheimer’s Association (NIA-AA) suggestion, MCI is a clinical-pathologic entity, i.e., symptoms/signs, rather than a biological definition \cite{jack2018nia}. The symptom, such as the cognitive impairment and neurodegeneration, is a clinical consequence that may due to AD, other disease or their combination. This will lead to a high heterogeneity in MCI subjects and recruiting them without accessing their risks hierarchically may increase the failure probability of clinical trials, which might be the reason that AD research has made no progress for decades \cite{anderson2019state, libon2010heterogeneity}. Therefore, stratifying MCI patients into distinct risk subtypes owing to the etiology of AD or finally developing to AD is an important and urgent task, with the potential to tailor treatment for appropriate individuals and reduce the overall incidence of AD.


	
There have been many works aiming to classify MCIs into different subtypes. These methods stratify MCIs based on either the symptom severity or the progression status. The symptom-based methods primarily assess the severity of neurodegeneration via the cognitive test performance or neuroimaging biomarker profiles, with early MCI (EMCI) and late MCI (LMCI) as the well-known stratification paradigm \cite{aisen2010clinical, edmonds2019early}. A more refined method considered the cognitive impairment into multiple distinct domains, and the severity of each domain can jointly define the subgroup of MCIs \cite{winblad2004mild, brambati2010single, mukherjee2020genetic, hughes2011should}. The severity of neuroimaging-derived phenptype (IDP) and fluid biomarkers were also used as the divided principle. Nettiksimmons \emph{et. al.} \cite{nettiksimmons2014biological} perform clustering on magnetic resonance imaging (MRI) quantitative traits (QTs), cerebrospinal fluid concentrations (CSF), and serum biomarkers and identified four subtypes with different degrees of severity. Yang \emph{et. al.} \cite{yang2024gene} integrated sMRI with genetic variations to stratify patients, but it aimed for AD/MCI subtypes in combination other than MCI subtypes.


The progression-based methods classified MCIs into distinct groups depending on whether a MCI subject progress to AD or his/her neuroimaging QTs change significantly. Dong \emph{et. al.} \cite{dong2017heterogeneity} identified four distinct subtypes according to the atrophy patterns of longitudinal MRI scans. SuStaIn, a recently proposed staging subtyping method, yielded two distinct AD subtypes and accessed their respective progression sequences from health control to dementia \cite{baumeister2022two, baumeister2024generalizable}. Edmonds \emph{et. al.} \cite{edmonds2024data} performed clustering on neuropsychological data and identified three MCI subtypes with different progression profiles. Caminiti \emph{et. al.} \cite{caminiti2024fdg} employed longitudinal FDG-PET, which measures the hypometabolism, to classify MCI into three subtypes with different levels of hypometabolism. In addition, many works classified MCIs into stable MCI (sMCI) and progressive MCI (pMCI, developing to AD finally) based on the progression status \cite{petrella2011default, prestia2013prediction}. Although the above approaches have gained success in subtyping MCIs or ADs into distinct subgroups, limitations exist. First, the dementia syndrome is not a specific biological concept, and thus the symptom-based methods are somewhat superficial since many symptoms of MCIs are not specific for AD. For example, the neurodegeneration and CSF total tau (T-tau) could derive from a variety of etiologies, with AD being only one of them \cite{jack2018nia}. Second, the progression-based methods is consequentialism and an individual's final disease state cannot be determined only if the follow-up clinical data is collected \cite{tong2016novel}. In terms of the purpose of early intervention, both aforementioned types of subtyping methods fail to figure out MCI patients of higher risk as early as possible, because they cannot well stratify the risk of MCIs or commit to other goals.

\begin{figure*}[htbp]
	\centering
	\includegraphics[width = \linewidth]{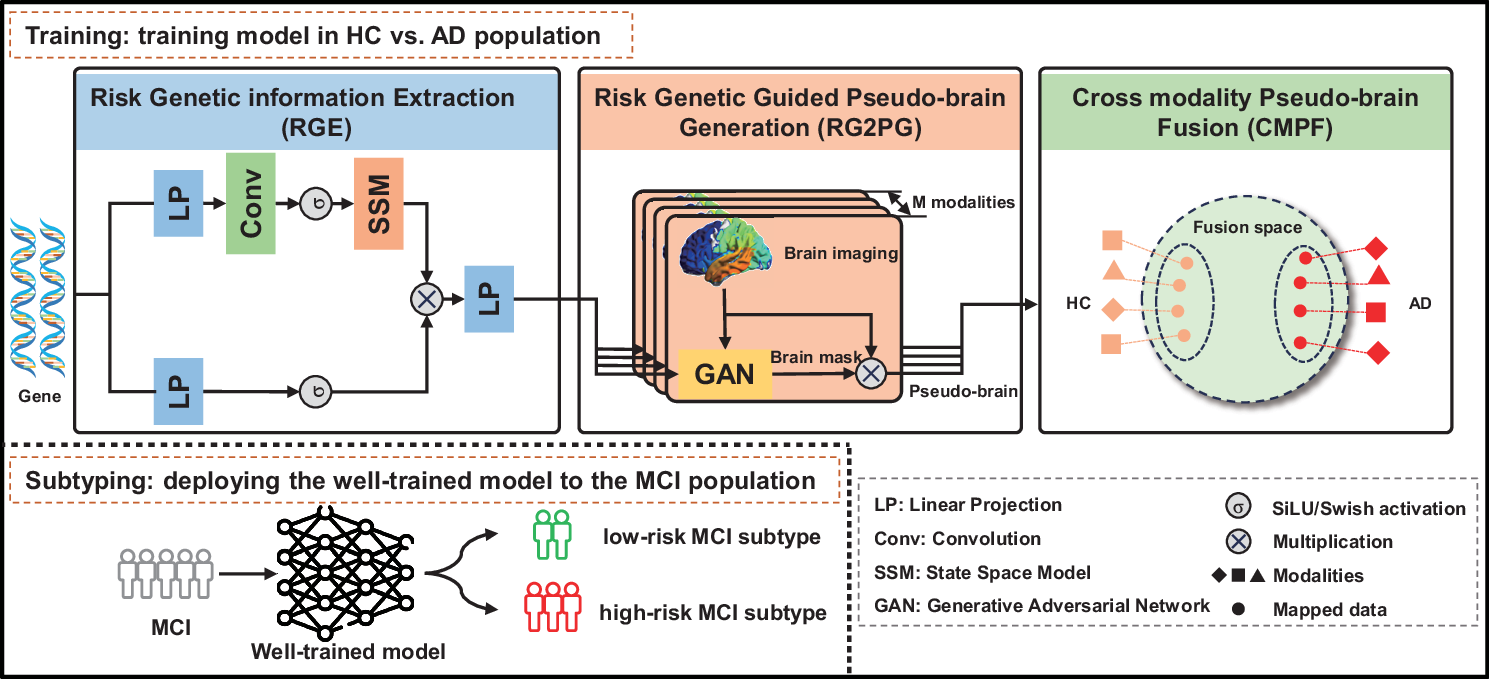}
	\caption{The framework of the proposed BigFirst method. The model comprises three modules: (1) the Risk Genetic information Extraction (RGE) module, which extracts disease-related genetic representations from extensive genetic loci; (2) the Risk Genetic Guided Pseudo-brain Generation (RG2PG) module, which leverages the genetic representations from RGE to generate multimodal disease-specific pseudo-brain imaging data; and (3) the Cross Modality Pseudo-brain Fusion (CMPF) module, which integrates the multiple pseudo-brain modalities using a multimodal fusion approach for disease prediction and MCI risk subtype identification. These three modules are connected in series to identify subtypes.} \label{DGMBF}
\end{figure*}

To better delay the progression of MCIs and decrease the incidence of dementia due to AD, we aim to screen out appropriate MCI subjects to undergo clinical trails. Considering the complicated pathophysiologic and pathogenic mechanism of MCIs, we here propose a novel Brain Imaging Genetics Fusion method to Identify MCI Risk SubTypes (BigFirst) to accommodate neuroimaging phenotypes and genetic variations. BigFirst comprises three modules, taking into consideration both the innate genetic information and acquired phenotypic information. The risk genetic information extraction (RGE) module extracts disease-related genetic representations from the whole-genome. The risk genetic guided pseudo-brain generation (RG2PG) module leverages the genetic representations to yield lesional brain feature maps for multiple modalities, including structural MRI and positron emission tomography (PET). Then lesional feature maps are fused with multiple brain imaging modalities to yield brain imaging lesion features for each imaging modality. The cross-modality pseudo-brain fusion (CMPF) combines these lesion brain imaging features for the disease prediction and MCI risk subtyping. The BigFirst with the three modules is illustrated in Fig.\ref{DGMBF}. We applied BigFirst to the Alzheimer’s Disease Neuroimaging Initiative (ADNI) dataset and stratify MCIs into two subgroups with different level risks. We further compared the two MCIs subtypes by investigating the genetics, demographics, fluid biomarkers, brain imaging features, and cognitive function at baseline, as well as their longitudinal developmental trajectories, because of the MCI symptoms can be manifested by abnormal phenotypes including brain imaging QTs and fluid  biomarkers \cite{dong2017heterogeneity,nettiksimmons2014biological, edmonds2016heterogeneous}, and could be caused by mutated genetic variations \cite{wen2022genetic}, unhealthy lifestyle factors \cite{katayama2021lifestyle}, and the comorbidity \cite{katabathula2023comorbidity, vassilaki2015multimorbidity}. Additionally, we identified potential lifestyle factors that may increase the risk of progression, providing valuable insights for patient monitoring and risk stratification.

\section{Results}\label{Results}
\subsection{BigFirst: a brain imaging genetics fusion method to stratify MCIs into different level of risks}\label{result_bigfirst}
BigFirst aims to identify MCI risk subtypes to select appropriate patients to undergo intervention as early as possible. Thus it only requires the cross-sectional brain imaging data and does not depend on the follow-up data. To better reveal the essence of high-risk populations, we also extract the causative genetic variants from the whole-genome to combine with brain imaging QTs exhibiting abnormalities in structure and function. BigFirst is trained on HC and AD subjects to ensure that the differences between MCI subtypes are disease-related. This learning strategy can access the risk level of MCIs and help filter out individuals with high or low risk. Specifically, we used Florbetapir-PET (AV45-PET) scans, fluorodeoxyglucose PET (FDG-PET) scans and volume-based morphometry MRI (VBM-MRI) scans with aim to capture the diversity abnormality across multiple brain imaging modalities, including the amyloid deposition, brain glucose metabolism rate and neurodegeneration due to the brain atrophy \cite{ossenkoppele2015atrophy, chew2013fdg, hardy2003relationship}. We employ 359,997 brain tissue related single nucleotide polymorphisms (SNPs), which were pre-selected from the whole genome using the GTEx version 8 database (See Section ~\ref{ADNI} for more details). We included 648 samples including 131 healthy controls (HCs), 100 ADs, and 417 MCIs. The detailed demographic characteristics were shown in the Table~\ref{Tab_demo}.
\begin{table}[h]
	\caption{Participant characteristics.}\label{Tab_demo}%
	\begin{tabular}{@{}llll@{}}
		\toprule
		&HC & MCI & AD \\
		\midrule
		Num(n)    					&	131 & 417     &100  \\
		Age(mean $\pm$ std)    		&	73.37$\pm$5.85  & 71.03$\pm$7.21   &74.08$\pm$8.58   \\
		Gender(M/F)    				&	61/70  & 201/216   &57/43   \\
		Education(mean $\pm$ std)   &	16.53$\pm$2.59  & 16.28$\pm$2.67    &15.51$\pm$2.68    \\
		Hand(R/L)    				&	122/9  	& 366/51   &91/9  \\
		$\emph{APOE}$ (N/P/NAN)   &	92/38/1  & 234/182/1  &32/68/0  \\
		\botrule
	\end{tabular}
\end{table}

BigFirst was first trained on HCs and ADs populations and the classification performance of HC. vs. AD was shown in Supplementary ~\ref{Disease prediction and ablation study}. Then the well-trained model was deployed on 417 MCI patients and yielded 86 low-risk MCIs and 331 high-risk MCIs. The algorithmic evaluation were first accessed and the differences between the two MCI subgroups were also evaluated via statistical analysis on cross-sectional and longitudinal data including phenotypic, genetic, cognitive, and environmental factors.
	
In phenotypic analysis, we primarily investigated brain imaging QTs and fluid biomarkers. The differences between imaging QTs (derived from AV45-PET, FDG-PET and VBM-MRI), CSF A$\beta$, plasma phosphorylated tau (p-Tau) and plasma neurofilament light (NfL) concentration levels of two subgroups were compared via the $t$-test with Bonferroni correction for multiple testing if needed \cite{weisstein2004bonferroni}. Given the established role of hippocampus in AD, we also examined the differences of hippocampus subfields such as CA1, CA2-3, CA4-DG, fimbria, hippocampal fissure, hippocampus, presubiculum and subiculum \cite{iglesias2015computational}. 

In the cognitive analysis, we assessed four domains to accommodate the memory, executive, language and visuospatial domains via the $t$-test. The difference of 28 clinical symptom-related environmental factors were also compared between two MCI subgroups via the $\chi^2$-test. The 28 clinical symptom-related environmental factors were described in Section ~\ref{Other data}. We also analyzed the differences of five covariates, including three discontinuous covariates (gender, handedness and $APOE$ genotype) with $t$-test and two continuous covariates (age and years of education) with $\chi^2$-test. To investigate whether the progression patterns show substantive difference between low-risk MCIs and high-risk MCIs, we conducted a longitudinal comparison using follow-up data collected at baseline (BL), 12-month (12M) and 24-month (24M) time-points. The development trajectories fitted from the mean values of longitudinal biomarkers were employed. To test whether genetic variations take charge of the difference for low-risk MCIs and high-risk MCIs, we conducted the genome-wide association study (GWAS) with the above five covariates included using PLINK version 1.9 \cite{chang2015second}. The significance threshold was set to $p =  1\times 10^{-5}$ since no SNP passed the initial $1\times 10^{-8}$ test. To further validate the biological relevance of the identified loci at the gene level, we performed gene-set enrichment analysis (GSEA) using MAGMA software \cite{de2015magma}. Moreover, we additionally performed the phenotype-wide association analysis (PheWAS) using publicly available data, including 4756 GWAS summay statistics, from the GWAS Atlas32 (https://atlas.ctglab.nl). The Bonferroni corrected significance threshold ($p<1.05\times 10^{-5}$) was used for PheWAS analysis, and this could help reveal related traits influencing the risk of AD.

\subsubsection{BigFirst identified two MCI subtypes}\label{Clustering performance}
BigFirst identified two MCI subtypes with different levels of risk. The clustering performance respect to the two MCI subtypes was evaluated. We used the Silhouette Coefficient (SI), Calinski-Harabasz Index (CH), and Davies-Bouldin Index (DB) as the evaluation criteria because no ground truth labels were provided. SI measures the coherence of identified clusters, and higher scores indicate more coherent clusters, i.e., better clustering performance. CH is the ratio of the sum of inter-cluster dispersion and the sum of intra-cluster dispersion, with a higher CH indicating a better clustering. DB is calculated the similarity between every cluster and its neighbour cluster, and smaller values stand for better clustering performance. We conducted comparative analyses between four respective clustering algorithms such as k-means, spectral clustering (SC), Gaussian Mixture Model (GMM) based clustering and deep contrastive learning-based clustering \cite{li2021contrastive}. Additionally, a reduced version of BigFirst with REG module removed was also used for comparison. The clustering performance was presented in Table~\ref{cluster_performance}. BigFirst, including the completed and reduced versions, obtained higher SI and CH values, with DB values being smaller than the remaining methods except for the SC method. Therefore, the proposed integrated learning framework outperformed existing clustering methods, showing a better performance in stratify MCI poputations.

\begin{table}[!htbp] 
	\caption{Clustering Performance on the ADNI data.}\label{cluster_performance}%
	\begin{tabular}{@{}lccc@{}}
		\hline
		&Silhouette Coefficient & Calinski-Harabasz Index & Davies-Bouldin Index \\
		\hline
		k-means	&	$0.15$    	&	$70.98$ 	& $2.26$ \\
		SC	&	$0.24$    	&	$2.45$ 	& $0.62$ \\
		GMM	&	$0.15$    	&	$70.98$ 	& $2.26$ \\
		CC	&	$0.36$    	&	$194.30$ 	& $1.08$ \\
		without RGE	&	$0.48$    	&	$420.38$ 	& $0.82$ \\
		with RGE	&	$0.52$    	&	$419.80$ 	& $0.74$ \\
		\hline
	\end{tabular}
\end{table}
	
\subsection{BigFirst can identify risk genetic variations from the whole-genome} \label{high_dim_gene}
As a by-product, BigFirst can extract risk genetic variations from the whole-genome genetic data during training. BigFirst used Mamba, a deep technique can recognize feature importance, to identify SNPs of great importance owing to its capability in handling high-dimensional sequential data \cite{gu2023mamba}. We performed a linear regression between the input (SNPs) and output of Mamba (representations of SNPs) and ranked SNPs according to their weights. Based on the weights, we identified loci situated in \emph{ST7L}, \emph{CAPZA1}, \emph{FAM13A}, \emph{FAM13A-AS1} and \emph{NQO2} genes. Among them, \emph{ST7L} and \emph{CAPZA1} was documented to be responsible for the hippocampus and the nucleus accumbens. The $\emph{ST7L}$ gene may play a role in AD by modulating the Wnt signalling pathway \cite{ng2019wnt, tabnak2021regulatory}. Moreover, besides AD, $\emph{CAPZA}$ showed association with other neurodegenerative disorders \cite{tuerk2017genetic, zhang2024heterozygous, lovrecic2009gene}, demonstrating that the atrophy in hippocampus and the nucleus accumbens were nonspecific for AD. Loci of $\emph{FAM13A}$ were in the intron region or promoter region and have also shown statistically significant associations with Alzheimer's disease at the genome-wide level \cite{parrado2016p4}. \emph{NQO2}'s loci has also been shown to be associated with neurodegenerative diseases including AD \cite{janda2024polymorphisms, janda2015parkinsonian}. To further validate the capability of handling whole-genome data, we conducted ablation experiments, with the results presented in Supplementary ~\ref{Disease prediction and ablation study}.

	
\subsection{Investigation of baseline differences between MCI subtypes}\label{Subtype cross-sectional analysis}

	\begin{figure}[!htbp]
		\centering
		\begin{subfigure}{0.325\linewidth}
			\rotatebox{90}{\scriptsize{low-risk vs. high-risk}}
			\centering
			\includegraphics[width=0.9\linewidth]{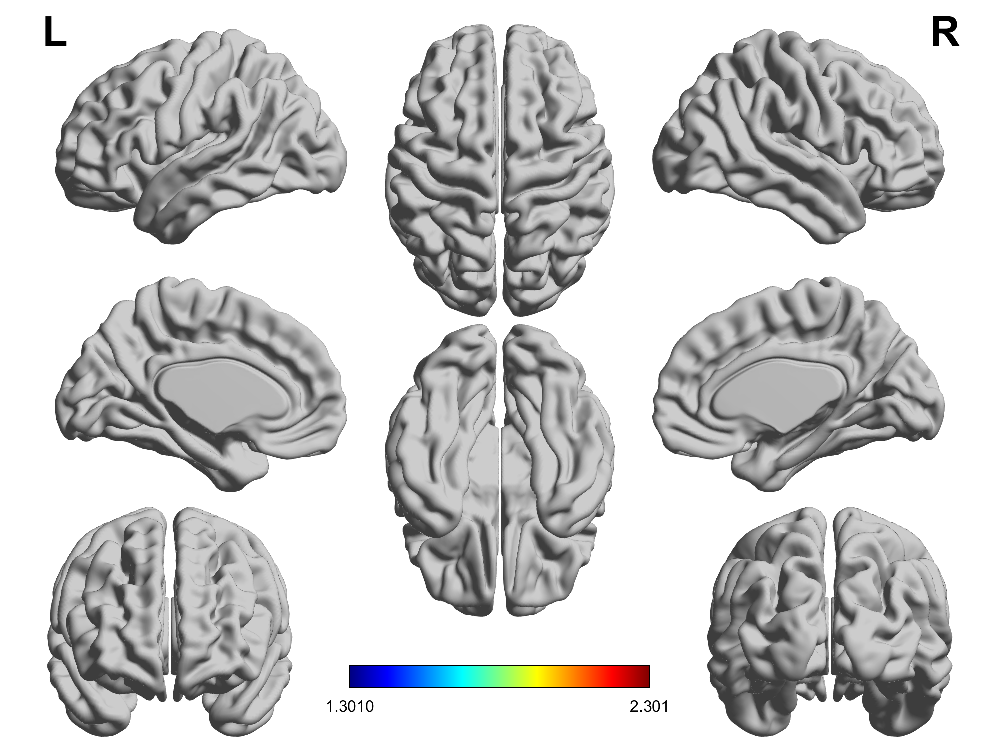}
			\caption{AV45}
			\label{Fig_MCI_AV45_BL}
		\end{subfigure}
		\centering
		\begin{subfigure}{0.325\linewidth}
			\centering
			\includegraphics[width=0.9\linewidth]{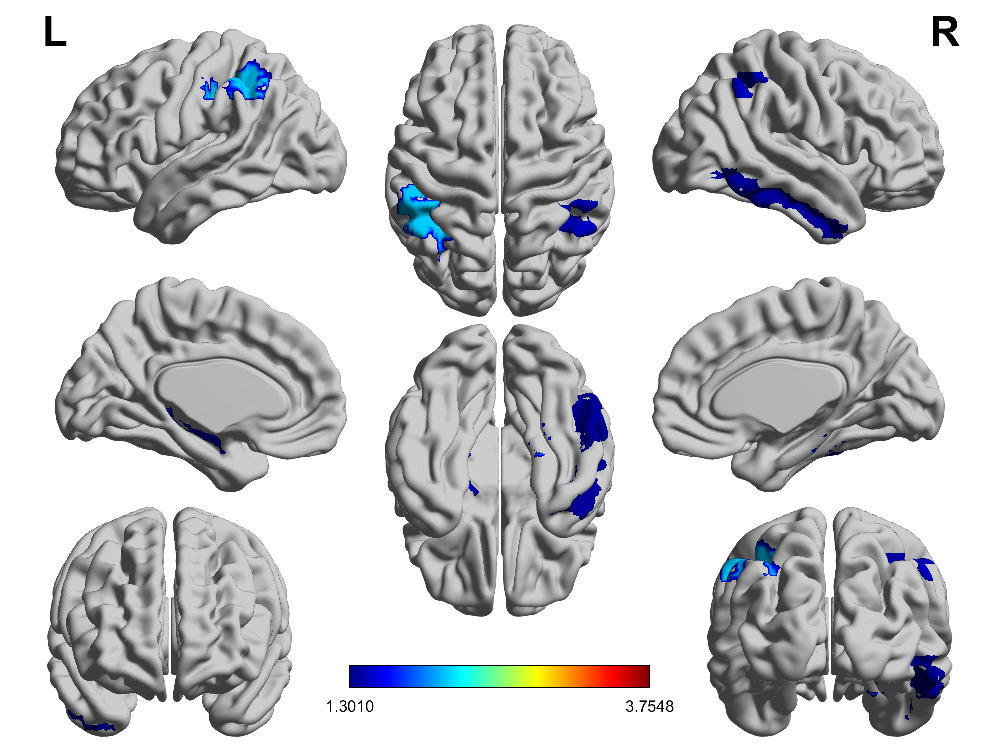}
			\caption{FDG}
			\label{Fig_MCI_FDG_BL}
		\end{subfigure}
		\centering
		\begin{subfigure}{0.325\linewidth}
			\centering
			\includegraphics[width=0.9\linewidth]{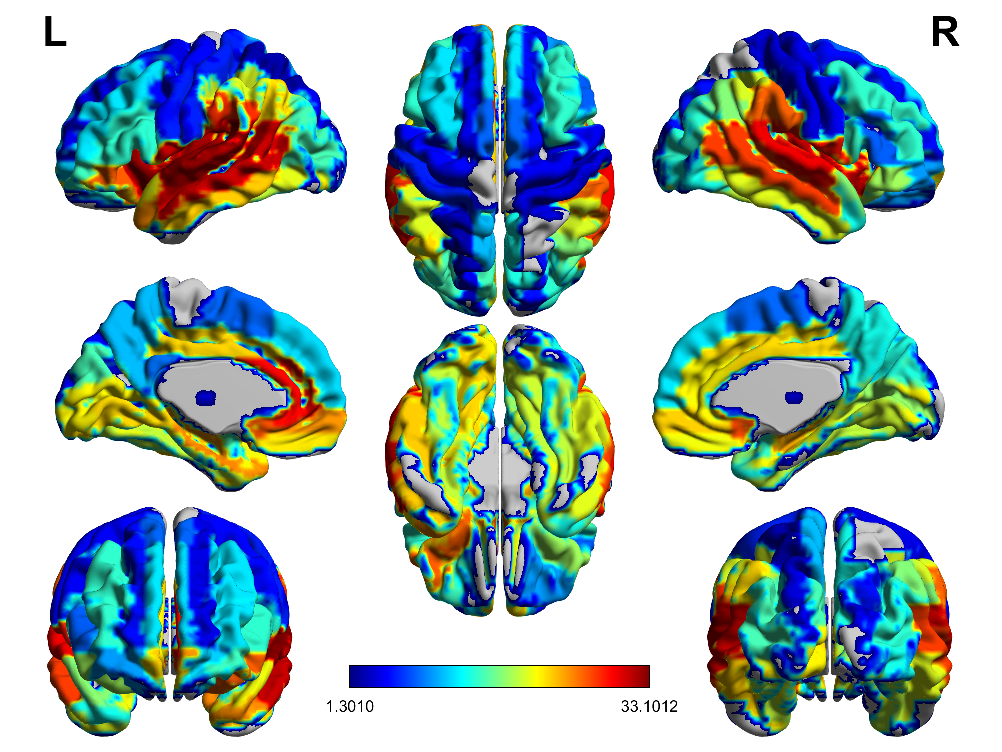}
			\caption{VBM}
			\label{Fig_MCI_VBM_BL}
		\end{subfigure}
		\newline
		\centering
		\begin{subfigure}{0.325\linewidth}
			\rotatebox{90}{\scriptsize{~~HC vs. low-risk}}
			\centering
			\includegraphics[width=0.9\linewidth]{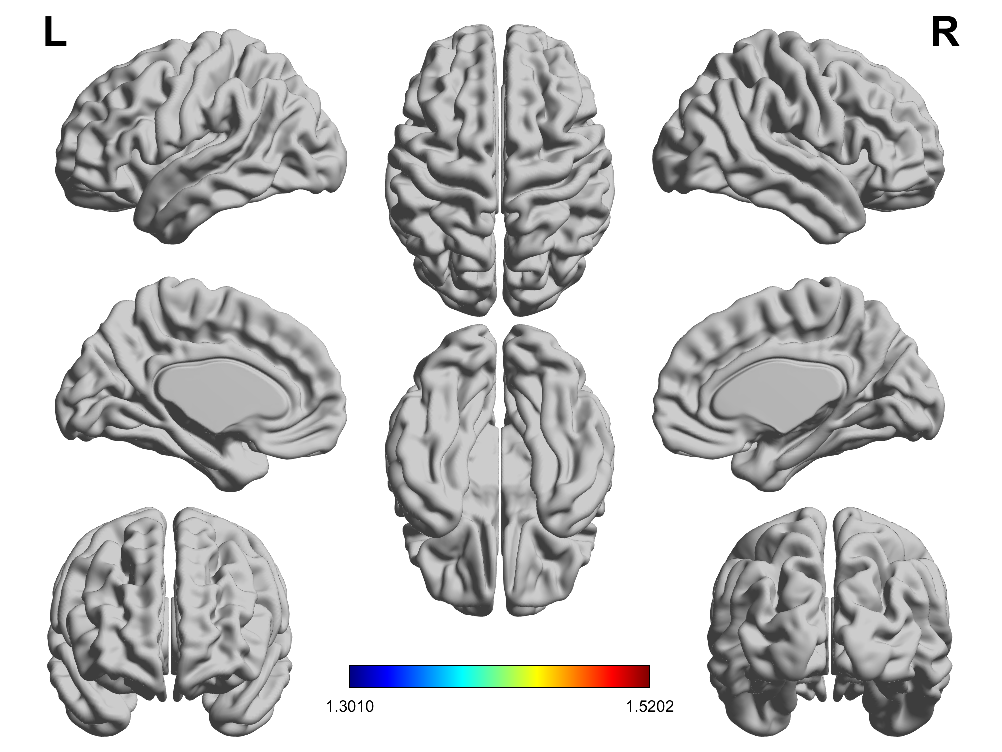}
			\caption{AV45}
			\label{Fig_HC_AV45_C}
		\end{subfigure}
		\centering
		\begin{subfigure}{0.325\linewidth}
			\centering
			\includegraphics[width=0.9\linewidth]{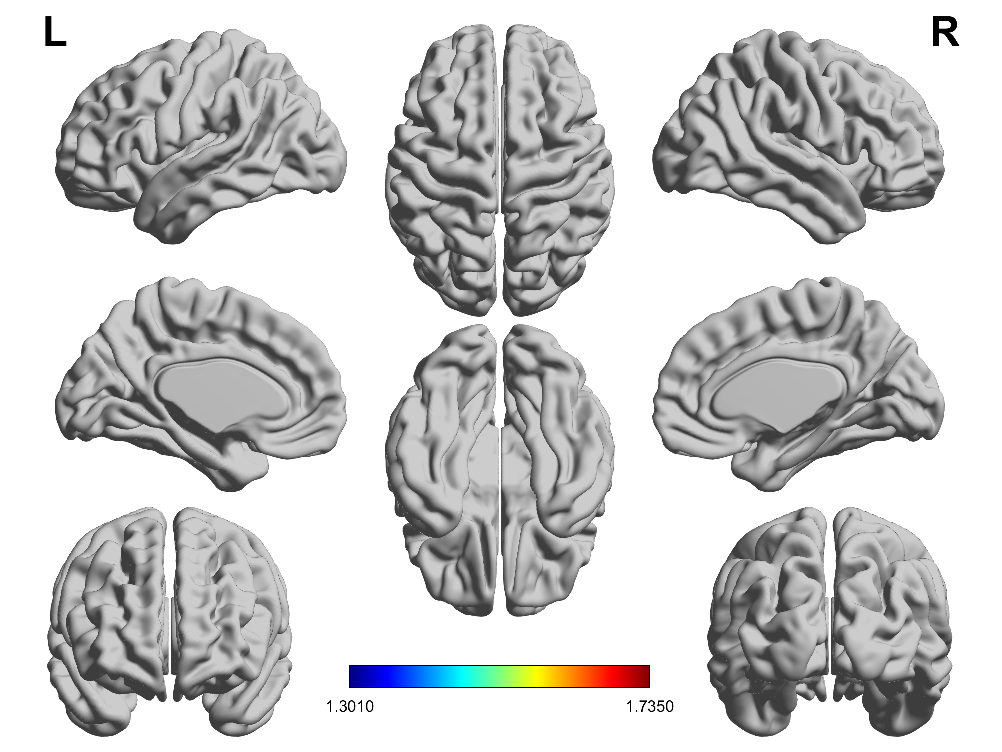}
			\caption{FDG}
			\label{Fig_HC_FDG_C}
		\end{subfigure}
		\centering
		\begin{subfigure}{0.325\linewidth}
			\centering
			\includegraphics[width=0.9\linewidth]{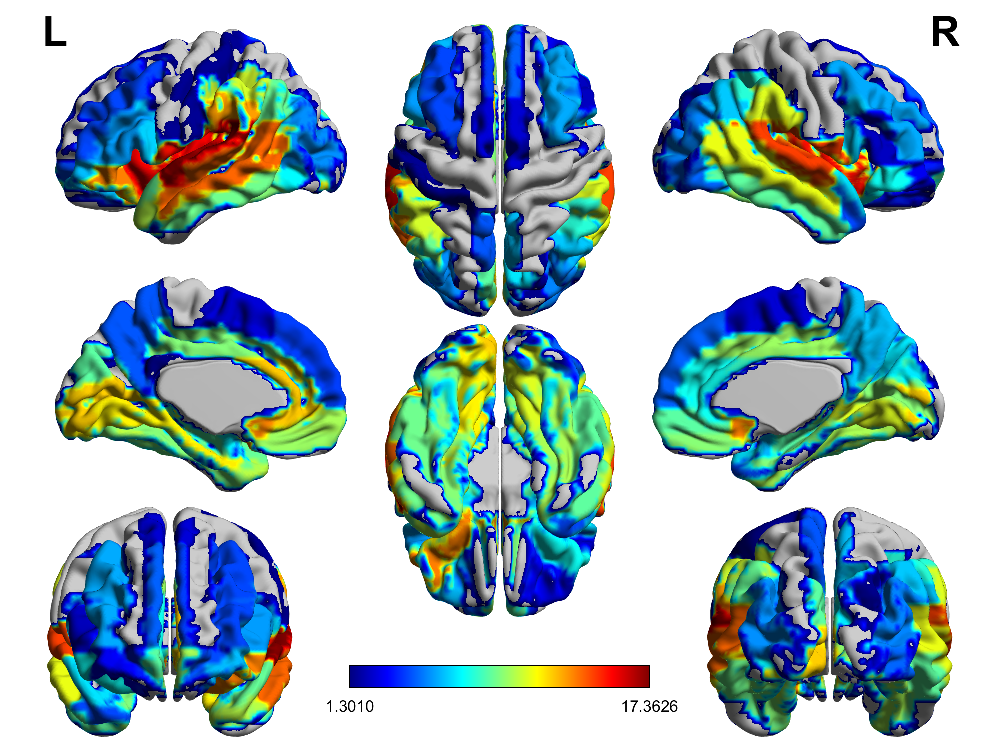}
			\caption{VBM}
			\label{Fig_HC_VBM_C}
		\end{subfigure}
		\newline
		\centering
		\begin{subfigure}{0.325\linewidth}
			\rotatebox{90}{\scriptsize{~~AD vs. high-risk}}
			\centering
			\includegraphics[width=0.9\linewidth]{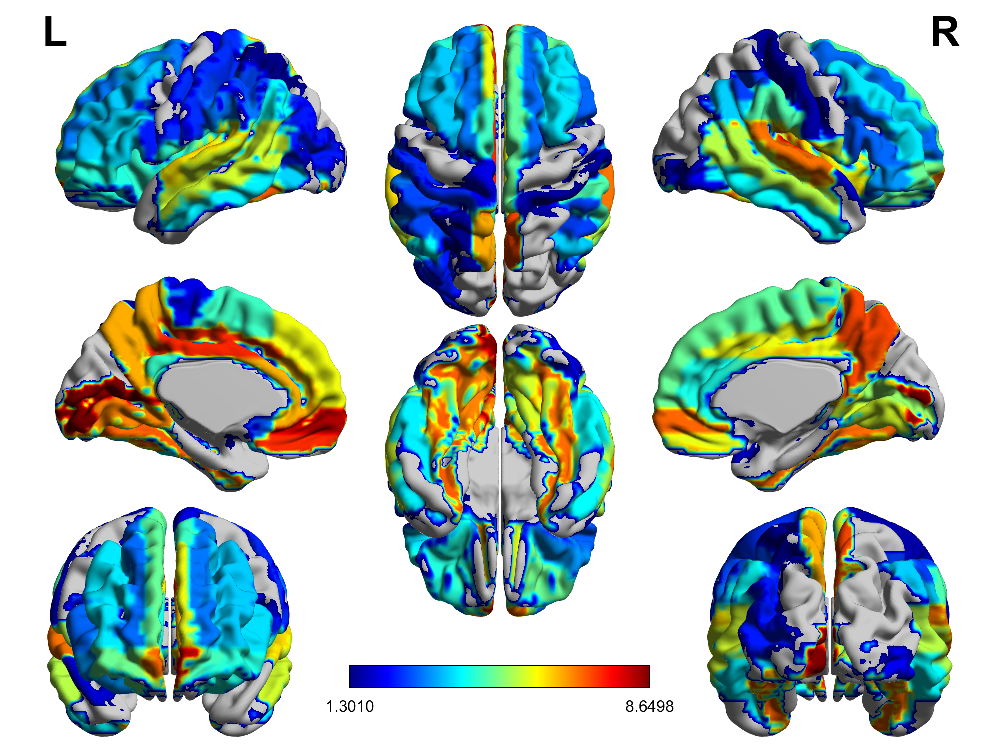}
			\caption{AV45}
			\label{Fig_AD_AV45_C}
		\end{subfigure}
		\centering
		\begin{subfigure}{0.325\linewidth}
			\centering
			\includegraphics[width=0.9\linewidth]{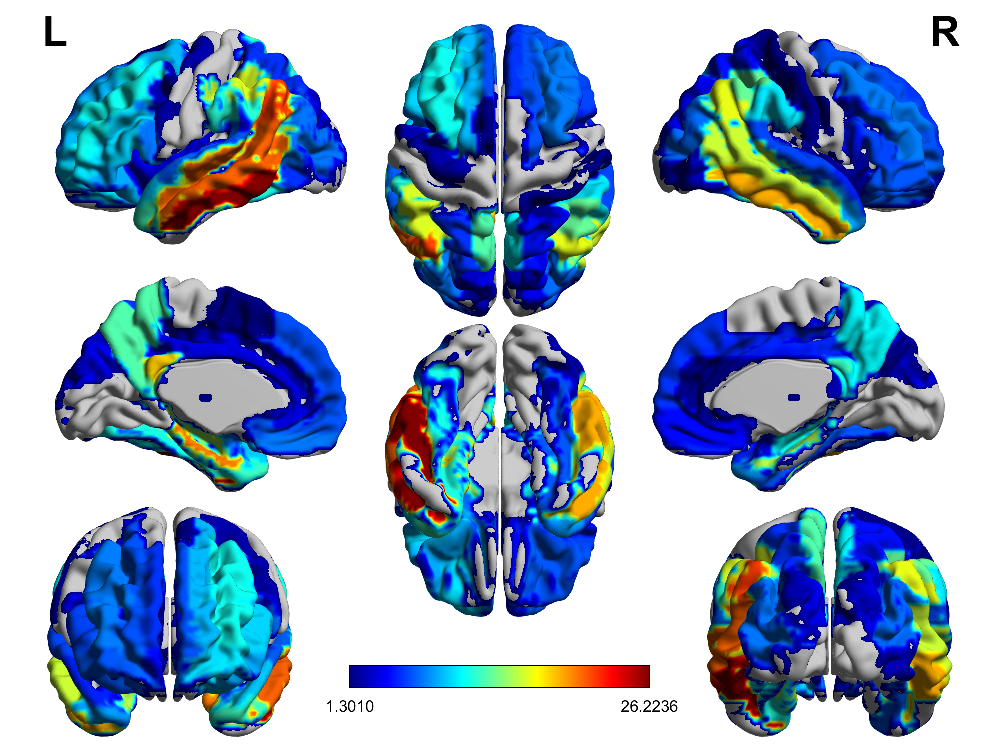}
			\caption{FDG}
			\label{Fig_AD_FDG_C}
		\end{subfigure}
		\centering
		\begin{subfigure}{0.325\linewidth}
			\centering
			\includegraphics[width=0.9\linewidth]{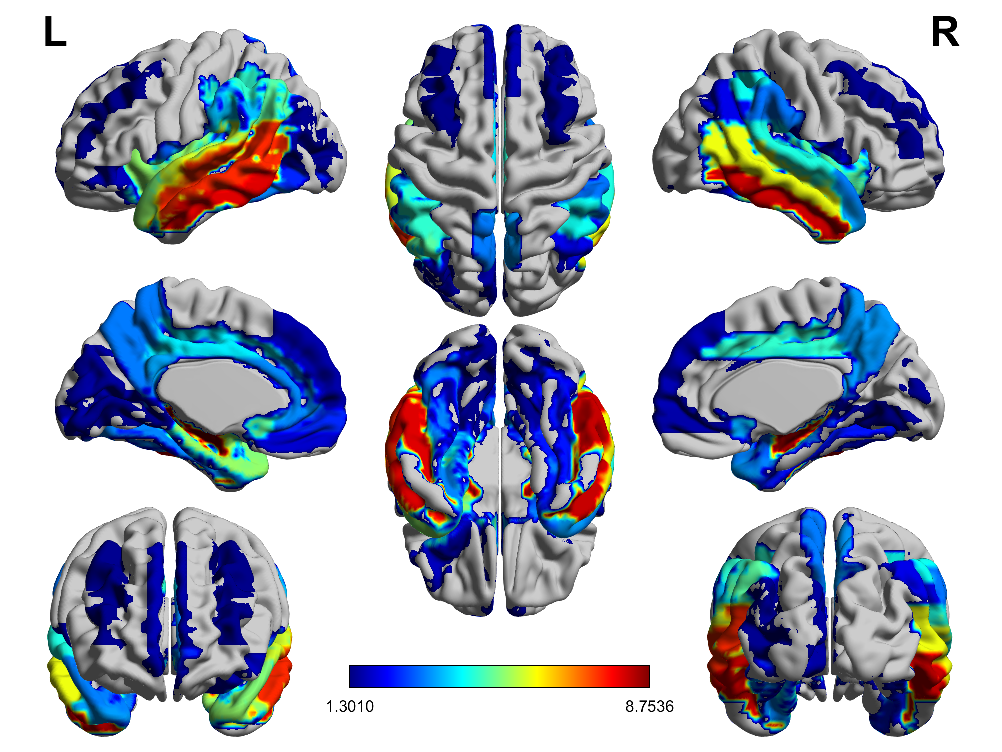}
			\caption{VBM}
			\label{Fig_AD_VBM_C}
		\end{subfigure}
		\caption{Differences in brain imaging QTs between (a-c): low-risk MCI and high-risk MCI, (d-f): HC and low-risk MCI, (g-i): AD and high-risk MCI. The gray area in the figure represents non-significant regions, while the colored areas indicate the -log($p$) values of significant brain regions. More intense red hues correspond to smaller $p$-values (higher significance), whereas bluer regions indicate larger $p$-values (lower significance). All statistically significant brain regions shown in the figure have passed Bonferroni correction. In low-risk vs. high-risk MCI and HC vs. low-risk MCI tasks, more extensive regional brain differences were observed in VBM compared to AV45 and FDG, suggesting that brain atrophy is a more sensitive discriminative biomarker and may occur earlier than changes in the other two modalities. However, in AD vs. high-risk MCI task, all three modalities showed widespread differences, indicating greater divergence between these two subject groups compared to others.}
		\label{Fig_QT_C}
	\end{figure}

\subsubsection{MCI subtypes show substantive differences in neuroimaging QTs}\label{QT_BL}
We investigated the differences between two MCI subtypes via testing brain imaging QTs derived from AV45-PET, FDG-PET, and VBM-MRI, using $t$-test with Bonferroni correction \cite{weisstein2004bonferroni}. The statistical results and significant different brain areas of each brain imaging modality were presented in Fig~\ref{Fig_QT_C}. After correction, the FDG-PET pattern showed prominent differences primarily in the hippocampus, parietal-inf, caudate, temporal lobe, cerebellum, and vermis, suggesting that the hypometabolism happening to these areas implicating a high risk of progression to AD. The patterns with respect to VBM-MRI scans showed widespread differences across nearly the entire brain, indicating significant variations in the degree of atrophy between the low-risk MCIs and high-risk MCIs. These findings suggest that patients with severe brain atrophy are at a higher risk, which aligns with existing knowledge and clinical experience. Furthermore, they imply that structural changes in the brain may precede metabolic alterations in the progression of AD. Given the hippocampus's established role area for AD, we also examined the differences of hippocampal subfields, including CA1, CA2-3, CA4-DG, fimbria, hippocampal fissure, hippocampus, presubiculum, and subiculum, between the two MCI subtypes \cite{iglesias2015computational}. Again, significant atrophy differences were observed in the CA2-3, CA4-DG, hippocampus, presubiculum, and subiculum subregions after Bonferroni correction, with these findings consistent across the left and right cerebral hemispheres. However, there was no brain area reached the significance level in terms of the AV45-PET QTs, indicating that the aggregation of amyloid beta (A$\beta$) plaques between the two subtypes was similar. The reason might be that A$\beta$ is an early biomarker of AD, and during the late stage of AD progression, all MCI patients could suffer from elevated A$\beta$ plaques \cite{schindler2018cerebrospinal, hansson2018csf, lowe2024amyloid}. These findings necessitated using of multi-modal neuroimaging QTs in subtyping MCIs.
	
\subsubsection{MCI subtypes show substantive differences in A$\beta$, p-tau and NfL}\label{Fluid_BL}
\begin{figure}[!htbp]
	\centering
	\includegraphics[width = \linewidth]{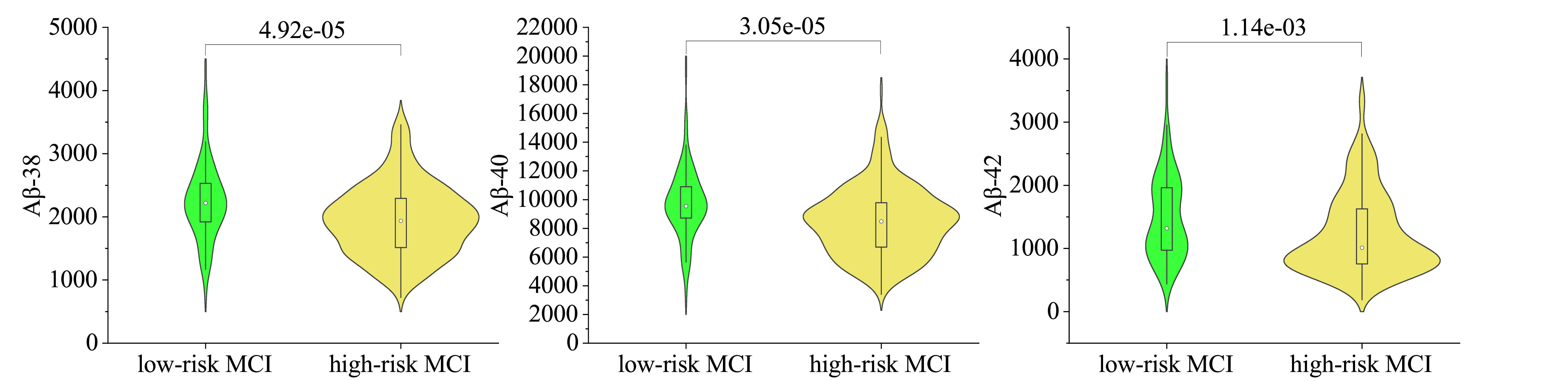}
	\caption{Differences in CSF A$\beta$ between two subtypes. Here we compared A$\beta$38, A$\beta$40, and A$\beta$42. Significant differences between the two subtypes were observed across all three proteins, with p-values labeled on the figure. The vertical axis represents the concentration (pg/mL) of each respective protein.} \label{Fig_Abeta_BL}
\end{figure}

We examined the differences between low-risk and high-risk MCI groups in terms of CSF amyloid beta levels (A$\beta$38, A$\beta$40, and A$\beta$42), plasma p-tau protein 181 (p-tau 181), and plasma NfL concentration levels at baseline using $t$-test. These biomarkers have been established as significant indicators of AD \cite{lewczuk2018plasma, mattsson2019association, thijssen2020diagnostic, andreasen2002beta}. There were significant differences in all three A$\beta$ biomarkers between two MCI subgroups (A$\beta$38: $p=4.92\times 10^{-5}$, A$\beta$40: $p=3.05\times 10^{-5}$, A$\beta$42: $1.14\times 10^{-3}$), with the distributions illustrated in Fig~\ref{Fig_Abeta_BL}. In contrast, no significant differences were observed in p-tau 181 ($p=3.57\times 10^{-1}$) and NfL levels ($p=7.90\times 10^{-2}$). These findings suggest that p-tau 181 and NfL were not as sensitive as A$\beta$ in distinguishing MCIs with different levels of risk.

\subsubsection{MCI subtypes show substantive differences in memory, executive function, language and visuospatial domains}\label{Cognition_BL}
\begin{figure}[!htbp]
		\centering
		\includegraphics[width = \linewidth]{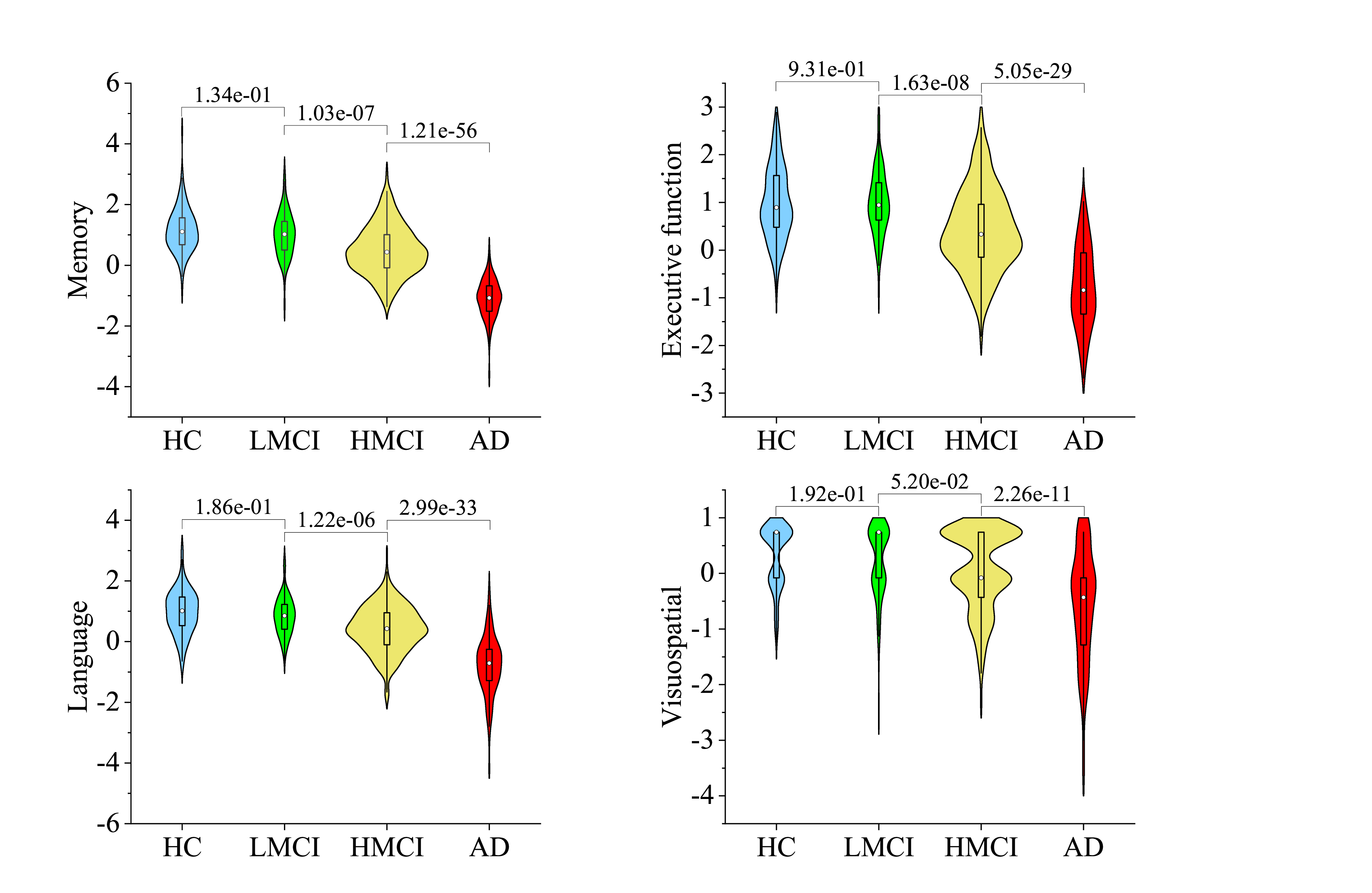}
		\caption{Differences between the cognitive domains of the four groups with $p$-values. LMCI: low-risk MCI, HMCI: high-risk MCI. The vertical axis represents the score of each cognition domain. The two subtypes exhibited significant differences across all three other cognitive domains, excluding visuospatial.} \label{Fig_CogDomain_BL}
\end{figure}

We here investigated the difference between MCI subgroups with respect to four AD-related domains, including the memory, executive function, language, and visuospatial \cite{weiler2014default, stopford2012working, mukherjee2020genetic}. To ensure early stratification, the baseline data was used. Significant differences were observed in three of the four domains, including the memory ($p=1.03\times 10^{-7}$), executive ($p=1.63\times 10^{-8}$) and language($p=1.22\times 10^{-6}$), with the visuospatial failing to reach the significance level ($p=5.20\times 10^{-2}$). These findings indicate a more pronounced deterioration in the memory, executive function, and language abilities in high-risk MCIs compared to low-risk ones, confirming previous findings. The comparison results were illustrated in Fig~\ref{Fig_CogDomain_BL}.

\subsubsection{MCI subtypes show substantive differences in clinical symptoms}\label{Clinical symptoms}
We analyzed 28 disease-related clinical symptoms, different from AD, in both subtypes of MCIs with different risks levels using the $\chi^2$-test. There were three symptoms reaching the significance level such as the headache ($p=6.50\times 10^{-3}$), urinary frequency ($p=2.08\times 10^{-2}$) and falling ($p=1.51 \times 10^{-2}$). According to previous studies, the headache was proved to be a risk factor to increase the conversion risk of MCI patients \cite{cermelli2024headache}, but the urinary frequency and falling were more likely to be the secondary symptoms caused by the cognitive impairment \cite{harlein2009fall, na2020relationship, ransmayr2008lower}.
	
\subsubsection{MCI subtypes show substantive differences in covariates}\label{Covariates}
We further investigated the differences of three discrete covariates, i.e., gender, hand, $\emph{APOE}$ genotype, via $\chi^2$-test. Another two continuous covariates including the age and education were also investigated via $t$-test. The results (see supplementary Fig~\ref{supp_COVS})  revealed a dominant difference on the gender, with males constituting a significantly higher proportion of high-risk MCI patients ($p=1.13 \times 10^{-2}$), suggesting a potentially higher AD incidence risk than females. The age also emerged as a significant factor ($p=1.97\times 10^{-9}$), with younger individuals exhibiting a lower risk. In contrast, the handedness, years of education, and $\emph{APOE}$ genotype showed no significant differences for the two MCI subgroups (hand: $p=9.78\times 10^{-2}$, edu: $p=4.52\times 10^{-1}$, $\emph{APOE}$: $p=4.10\times 10^{-1}$). The absence of significance of the $\emph{APOE}$ genotype may be explained by the fact that most of these MCI patients were derived by the mutated allele of $\emph{APOE}$, see supplementary Fig~\ref{supp_COVS}.
	
\subsection{Investigation of longitudinal differences between MCI subtypes}\label{Subtype longitudinal analysis}
\subsubsection{MCI subtypes show substantive progression difference in plasma tau and NfL}\label{Fluid_LT}
\begin{figure}[!htbp]
		\centering
		\includegraphics[width = \linewidth]{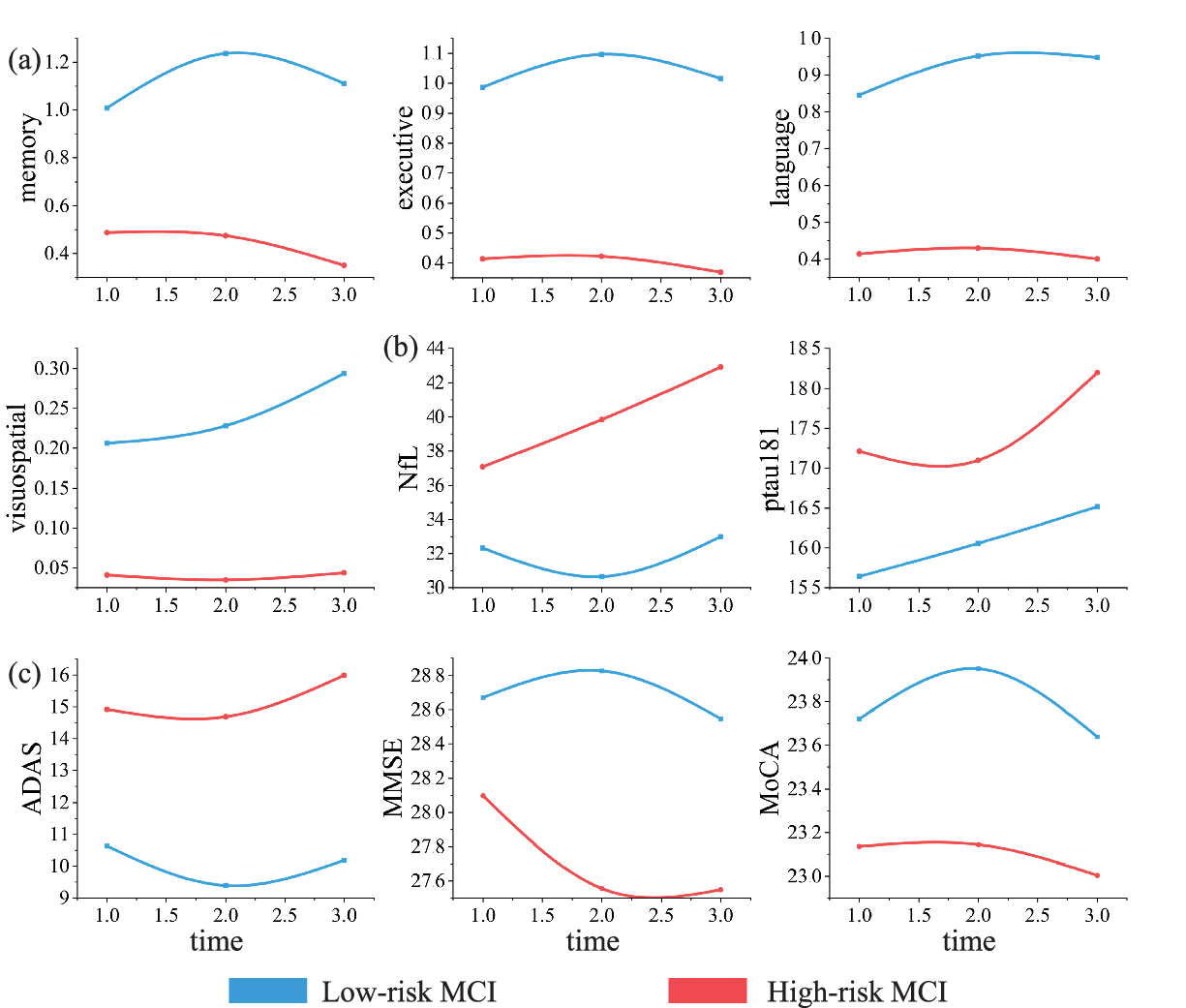}
		\caption{The longitudinal trajectories of the two subtypes in (a): four cognitive domain scores, (b): two fluid biomarkers, (c): three cognitive test scores. } \label{longitudinal}
	\end{figure}

We investigated the longitudinally progressing patterns of fluid biomarkers plasma p-tau181 and NfL concentrations between the two MCI subtypes. The longitudinal A$\beta$ data was limited and thus was ignored here. The longitudinal trajectories were illustrated in Fig.~\ref{longitudinal}. The results indicated that the progression of NfL was slower in the low-risk group compared to the high-risk group. A similar result was observed in the progression pattern of p-tau181, with the high-risk group exhibiting faster progression in later stages compared to the low-risk group.

\subsubsection{MCI subtypes show substantive progression differences in cognition ability}\label{Cognition_LT}
To investigate whether the cognitive progression of the two subtypes show difference, we conducted a longitudinal analysis of the memory, executive function, language and visuospatial domains. The progression trajectories for both subtypes were presented in Fig.~\ref{longitudinal} (a). 

The development trajectories of all four cognitive domains showed substantive difference, with the low-risk MCI group exhibiting slower progression compared to that of the high-risk MCIs. Interestingly, the low-risk MCI subtype exhibited an improvement over time. 
	
Additionally, given the widespread clinical use of the Alzheimer's Disease Assessment Scale (ADAS), Mini-Mental State Examination (MMSE), and Montreal Cognitive Assessment (MoCA) as diagnostic tools for AD, we conducted supplementary analyses to examine their longitudinal divergence between the two subtypes. The results consistently indicate that the low-risk subtype progresses more slowly than the high-risk subtype, especially in MMSE. The low-risk MCI subtype additionally exhibited indications of cognitive improvement during the early stages. The trajectories were illustrated in Fig.~\ref{longitudinal} (c).
	
\subsection{Genetic differences between MCI subtypes}\label{Genetic analysis}
\subsubsection{MCI subtypes have different genetic underpinnings}\label{GWAS}
\begin{figure}[!htbp]
		\centering
		\includegraphics[width = \linewidth]{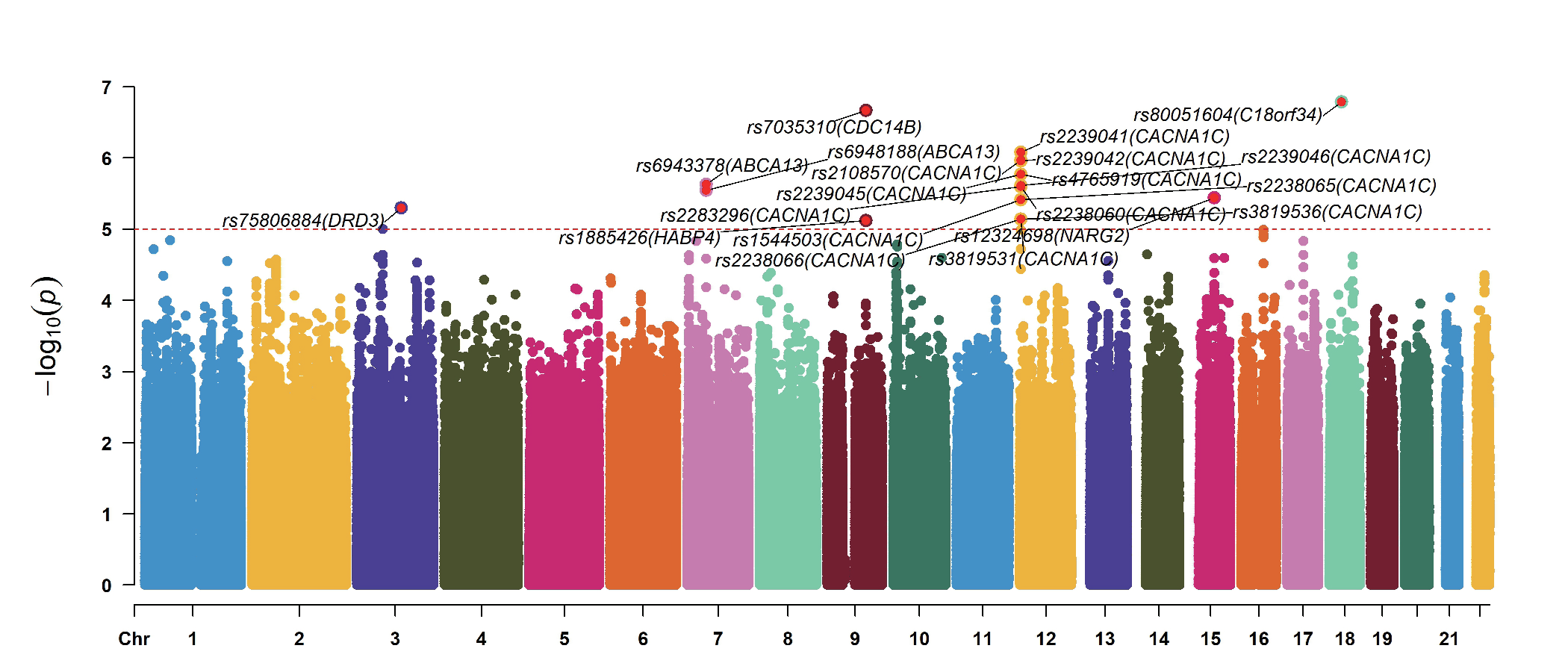}
		\caption{Genome-wide associations of risk subtypes. A genome-wide association study identified 20 genomic loci associated with MCI risk subtypes using a genome-wide significance threshold of $p=1 \times 10^{-5}$. The majority of loci were located within the $CACNA1C$ gene on chromosome 12.} \label{Fig_GWAS}
\end{figure}

We conducted a genome-wide association study (GWAS) to investigate the genetic basis underpinning the difference of the two MCI risk subtypes. The results were illustrated in Fig~\ref{Fig_GWAS}. The significant loci primarily located in the $\emph{CACNA1C}$ and $\emph{ABCA13}$ genes. The significant SNPs were presented in Table~\ref{Tab_GWAS}. Both $\emph{CACNA1C}$ and $\emph{ABCA13}$ genes have been demonstrated to be associated with various neurodegenerative disorders \cite{knight2009cytogenetic, moon2018cacna1c}. These SNPs were further investigated based on the gene-set analysis (GSEA). The GSEA revealed significant AD related gene sets, including $\emph{WU$\_$ALZHEIMER$\_$DISEASE$\_$UP}$ and $\emph{GOBP$\_$CELLULAR$\_$LIPID$\_$METABOLIC$\_$PROCESS}$. These findings suggested that $\emph{CACNA1C}$ and $\emph{ABCA13}$ may play a role in forming the difference risk for the two MCI subgroups. More results were detailed in the supplementary \ref{supp_GWAS}.

\subsubsection{PheWAS results of differential gene $\emph{CACNA1C}$}\label{PheWAS}
To test whether the two MCI subgroups were associated with different traits through comment genetic basis, we conducted the phenotype-wide association analysis (PheWAS). Based on the GWAS, we investigated both $\emph{ABCA13}$ and $\emph{CACNA1C}$. The PheWAS result revealed significant associations between MCIs subgroups and traits of psychiatric and cardiovascular. The details were showed in Fig.~\ref{Fig_Phewas_CACNA1C} and Table~\ref{Tab_GWAS}. However, the PheWAS result with respect to $\emph{ABCA13}$ did no pass the Bonferroni correction and was not presented. This findings suggested that the presence of comorbidities may increase the risk of patients with MCI. More results were detailed in Supplementary \ref{supp_GWAS}.
	
\subsection{MCI subtypes show high consistency with sMCI/pMCI subtyping but not the same}\label{conversion}
\begin{figure}[!htbp]
		\centering
		\includegraphics[width = \linewidth]{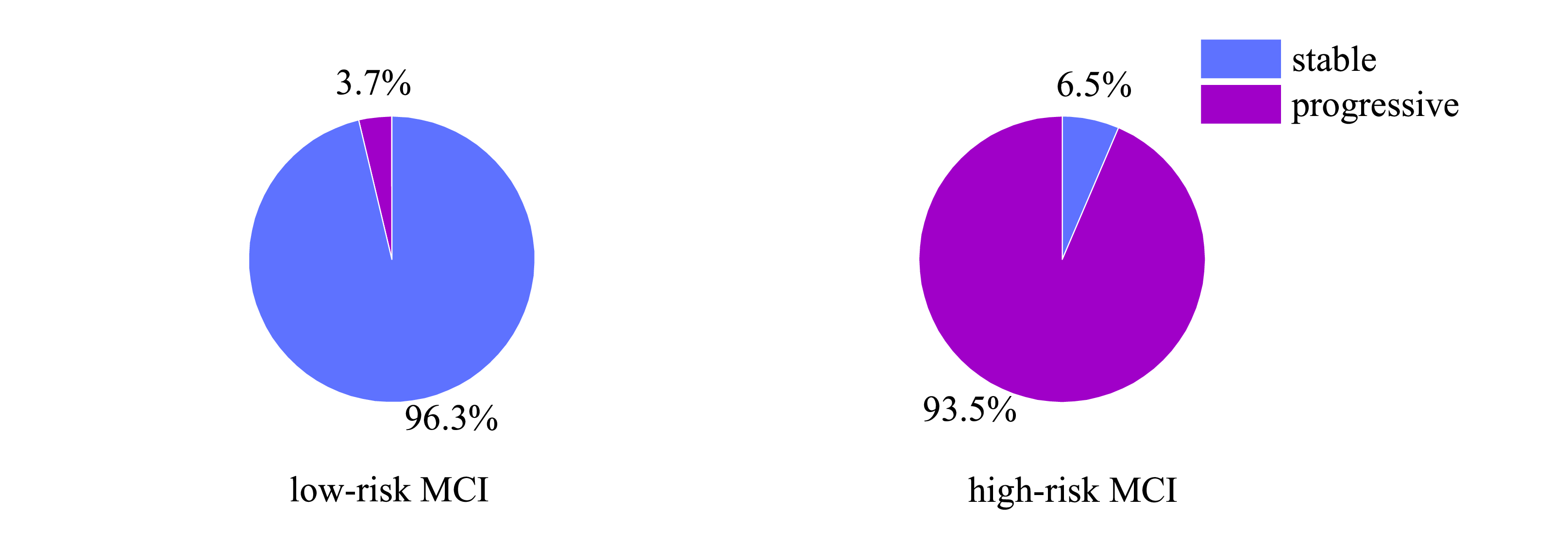}
		\caption{Differences in disease conversion between two subtypes over a two-year follow-up.} \label{Fig_Conversion}
\end{figure}

In this section, we investigated the difference between our subtyping methodology and the sMCI/pMCI subtyping methodology. As shown in Fig~\ref{Fig_Conversion}, our risk-based subtyping methodology showed substantial concordance with the two-year follow-up sMCI/pMCI classification methodology, demonstrating a strong ability in AD conversion prediction. However, it should be noted that low-risk designation did not implicate no conversion, and high-risk MCIs could also remain stable. The table status of the high-risk MCI patients might due to the short-term follow-up longitudinal data. This findings indicated that our subtyping methodology had a high consistency with the conventional sMCI/pMCI subtyping method, but not the same. Of note, our subtyping methodology only depends on the baseline neuroimaging data while the sMCI/pMCI subtyping requires follow-up data.

\subsection{Comparison of HC vs. low-risk MCI and high-risk MCI vs. AD}\label{Risk factors for disease stage conversion}
In this section, to further exploit the characteristics of the two MCI subgroups, we investigated the differences between HC and low-risk MCI groups, as well as that between high-risk MCIs and ADs.

\subsubsection{Comparison on AV45-PET, FDG-PET and VBM-MRI derived neuroimaging QTs}\label{QT_C}
Significant differences were observed between HCs and low-risk MCIs in the Cerebellum-3-Left region of AV45-PET, Vermis-8 of FDG-PET. The majority of brain regions on VBM-MRI showed substantive difference for HCs vs. low-risk MCIs. In contrast, individuals with high-risk MCI and AD exhibited significant differences across the majority of brain regions in all three imaging modalities, suggesting that high-risk MCI individuals may experience accelerated brain degeneration over a short period. The results were illustrated in Fig.~\ref{Fig_QT_C}.



\subsubsection{Comparison on CSF A$\beta$ 42, plasma tau, and NfL}\label{Fluid_C}
We examined the differences between HC and low-risk MCI subtypes in fluid biomarkers, as well as between high-risk MCI patients and AD patients. Significant differences were observed in CSF A$\beta$ 42 levels, plasma p-tau 181 levels, and NfL levels between the high-risk MCI subtype and AD groups after Bonferroni correction (A$\beta$42: $p=3.50\times 10^{-8}$, p-tau 181: $p=7.97\times 10^{-3}$, NfL: $p=6.27\times 10^{-4}$), but no significant differences were detected between HCs and the low-risk MCIs. These findings highlighted the importance of prioritizing A$\beta$42 , p-tau 181, and NfL as key biomarkers for assessing and monitoring the disease progression in the later stage of AD. 

\subsubsection{Comparison on cognition ability}\label{Cognition_C}
We compared cognitive scores across the aforementioned four domains. No significant differences were observed between HCs and low-risk MCI populations in any of the four domains. In contrast, significant differences were identified between high-risk MCI and AD populations. These results were illustrated in Fig~\ref{Fig_CogDomain_BL}.

\subsubsection{Comparison on covariates}\label{Covariates_C}
The five covariates of age, gender, handedness, years of education, and $\emph{APOE}$ genotype were compared between HCs vs. low-risk MCIs, and between high-risk MCIs and ADs. The $\emph{APOE}$ genotype and age exhibited significant differences in both comparison tasks (HC vs. low-risk MCI: $p_{APOE}=5.90\times 10^{-3}$, $p_{age}=1.99\times 10^{-13}$, AD vs. high-risk MCI: $p_{APOE}=9.44\times 10^{-6}$, $p_{age}=1.97\times 10^{-9}$). Although there was a significant difference in age between HC and low-risk MCI patients, it was the low-risk MCI patients who were significantly younger than the HC patients (Fig~\ref{supp_COVS}), implying that the disease accelerated cognitive deterioration in low-risk MCI patients. Significant differences in handedness were observed between HCs and low-risk MCIs ($p=1.52 \times 10^{-2}$), while the education level differed significantly between high-risk MCIs and ADs ($p=1.75 \times 10^{-2}$). 

\subsubsection{GWAS and PheWAS demonstrated significant lifestyle differences between HCs and two MCI subtypes}\label{PheWAS_C}
\begin{figure}[!htbp]
	\centering
	\includegraphics[width = \linewidth]{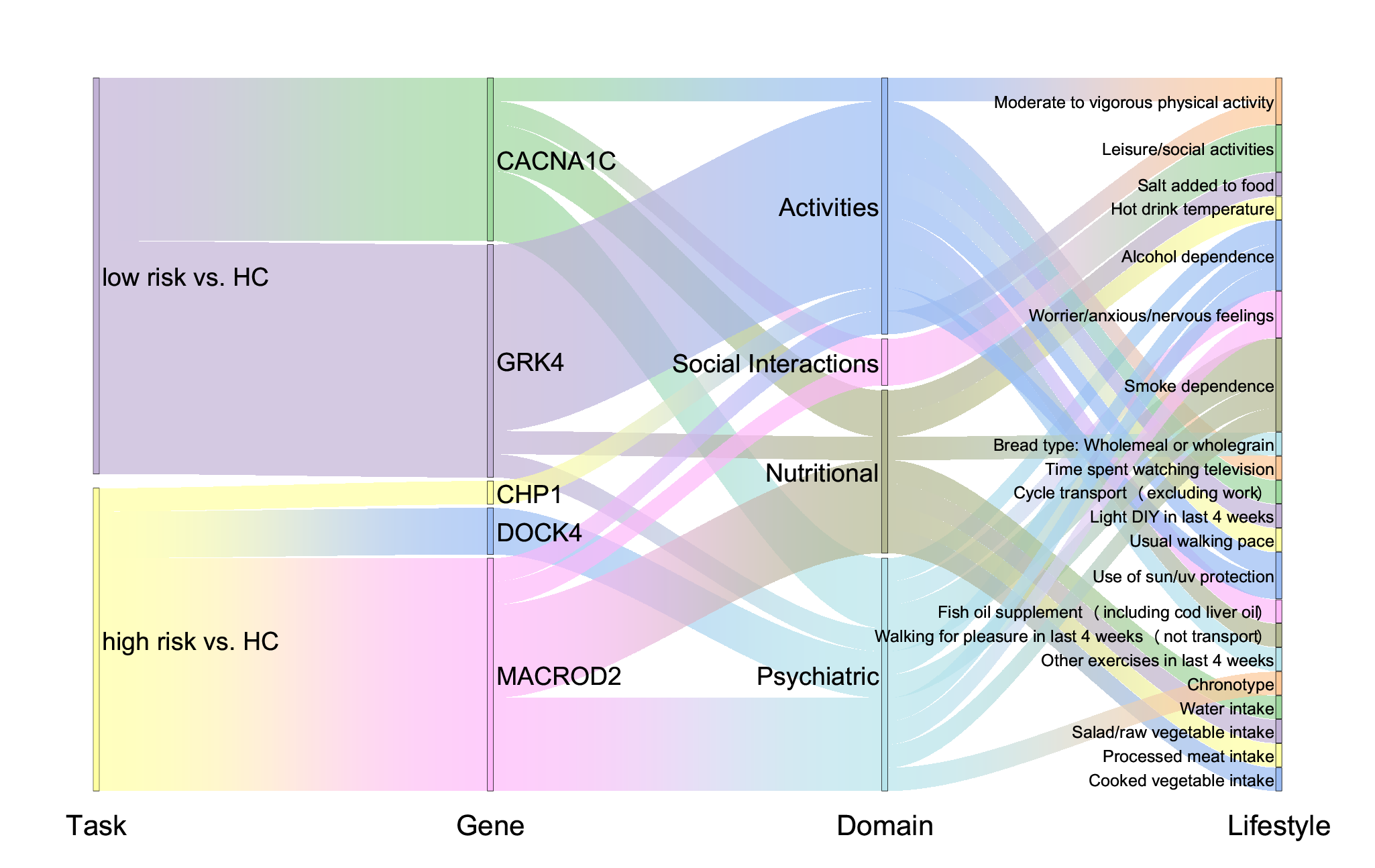}
	\caption{The relationships among tasks, genes, domains and lifestyles. In the visualization, the thickness of the connecting lines represents the -log($p$-value) from PheWAS results, with thicker lines indicating greater statistical significance. Both analyses identified associations with physical activity, social activity, dietary patterns, worrier/anxious/nervous feelings, and alcohol/tobacco dependency. Additionally, lifestyle differences between HC and low-risk groups involved transportation habits, while distinctions between HC and high-risk individuals were related to chronotype.} \label{Fig_sankey}
	\end{figure}

We first performed GWAS and then PheWAS to identify potential lifestyles or daily habits, underpinning via common genetic basis, between HCs vs. low-risk MCIs and HCs vs. high-risk MCI tasks. Consistent with the aforementioned analytical pipeline, we set the GWAS significance threshold to $1\times 10^{-5}$, and PheWAS with the Bonferroni-corrected threshold of $1.05\times 10^{-5}$. We performed PheWAS using genes reaching GWAS significance, and systematically extracted lifestyle-associated entries from the PheWAS results. Specifically, we identified the relationships among the HCs vs. low-risk MCIs and HCs vs. high-risk MCI tasks, genes identified by GWAS, PheWAS domains and traits. The results were shown in Fig.~\ref{Fig_sankey}. Both experiments identified that physical activity, social activity, dietary patterns, worrier/anxious/nervous feelings, and alcohol/tobacco dependency showed significant association with two MCI subtypes. The lifestyle differences between HCs and low-risk groups involved transportation habits, while chronotype showed prominent difference between HCs and high-risk MCI individuals.

\section{Discussion}\label{Discussion}
We proposed a brain imaging genetics fusion method to identify MCI risk subtypes (BigFirst) with the help of learning discriminative pattern between HC and AD. The method was grounded in the assumption that the physiological characteristics of low-risk MCI populations resembled those of HCs, whereas high-risk MCI populations were more akin to AD patients. This approach allowed us to obtain risk subtypes using cross-sectional neuroimaging data alone. Additionally, the follow-up investigations revealed that most of the biomarkers and disease-related phenotypes were significantly different between the two MCI subgroups, demonstrating the success of using our method in stratifying MCIs into subgroups with different levels of risk. The integrated learning strategy, accommodating both neuroimaging data and genetic information, enhanced the accuracy and specificity in understanding the heterogeneity of MCI populations, offering a more profound understanding of AD as well. Moreover, this was very meaningful and could help filter out appropriate individuals to undergo intervention or clinical trails.
	
	
After identifying two distinct MCI risk subtypes, a critical question raised: what physiological characteristics underlie the differences in risk? To answer this question, we conducted cross-sectional and longitudinal investigations in terms of genetic, phenotypic (brain imaging and fluid biomarkers), cognitive (cognitive function), and environmental (covariates and baseline symptoms) factors.
	
The GWAS, with significance level of $p=1\times 10^{-5}$, identified loci mainly located on two genes \emph{CACNA1C} and \emph{ABCA13}. This suggested that \emph{CACNA1C} and \emph{ABCA13} may affect the risk of transition of MCIs. The subsequent PheWAS using the two genes found that the difference between two MCI subgroups was significantly associated with some other brain disorders such as schizophrenia and bipolar disorder. This indicated that the presence of comorbidities may increase the risk of progression of MCIs.
	
We revealed that the two MCI subtypes differed significantly in FDG-PET and VBM-MRI scans, but not in AV45-PET neuroimaging data. The distribution of imaging QTs for VBM-MRI and FDG-PET data also differed. VBM-MRI showed significant differences in almost the whole brain, while the brain regions with significant differences in FDG-PET were concentrated in the hippocampus, parietal-infarct, caudate, temporal, cerebelum, and vermis. This demonstrated that AD was indeed strongly associated with cerebral atrophy relationship and, in addition to this, suggesting that brain atrophy may precede the onset of slowed glucose metabolism in the pathologic development of AD. Although the analysis on AV45-PET proved that there was no significant difference in A$\beta$ deposition between the two subtypes, there was still controversy about the relationship between A$\beta$-PET and CSF fluid A$\beta$ biomarkers for AD. Therefore, we further tested three A$\beta$ proteins such as A$\beta$38, A$\beta$40 and A$\beta$42, in addition to two plasma proteins including p-tau 181 and NfL concentrations. The results revealed that all three A$\beta$ biomarkers showed significant differences between two MCI subgroups, while the two plasma proteins did not. This demonstrated that the CSF fluid biomarkers A$\beta$ may be more sensitive than AV45-PET scans for distinguishing the two MCI risk subtypes. Although the cross-sectional concentrations of plasma biomarkers cannot be used independently to differentiate between the two subtypes, the progression rate of the low-risk MCIs was slower than that of the high-risk subtype, evidenced by the longitudinal analysis.
	
The cognition ability of the two MCI subgroups exhibited significant difference too. Three out of four cognitive domain (memory, executive, language, and visuospatial) showed substantive difference for the two MCI subgroups, and significant difference were also observed in ADAS, MMSE, and MoCA scores at baseline and longitudinal data. 
	
We also identified some environmental factors such as covariates and disease-related symptoms showing difference for two MCI subtypes. We found that older MCI patients have a higher risk, and male MCI patients had a higher probability of high risk than female patients. There was no significant difference for handedness, years of education and $\emph{APOE}$ genotype. This is interesting since that although $\emph{APOE}$ is a AD-risk gene, it cannot help stratify MCIs into different levels of risk. Therefore, $\emph{APOE}$ might not be served as a biomarker to filter out suitable MCI patients. We also identified three potentially relevant symptoms such as the headache, urinary frequency, and falls. By synthesizing insights from existing research, we speculated that the headache could serves as a predictor of stratifying risk levels, whereas urinary frequency and falls were by-products of the condition of MCI due to AD.
	
We further examined the relationship between our subtyping methodology and conventional sMCI/pMCI subtyping methodology. There was a high consistency between our subtyping methodology and that of sMCI/pMCI methodology. It is worth noting that low risk does not mean that it will not progress, and high risk does not necessarily mean that it will progress, which explains the observed discrepancies between our methodology and conventional sMCI/pMCI methodology. Another advantage of our subtyping methodology is that BigFirst does not depend on the longitudinal data which is very helpful for early intervention.
	
In addition to the difference investigation between MCI subtypes, we also tried to find more factors affecting during AD's progression spectrum, we compared HCs vs. low-risk MCIs and high-risk MCIs vs. ADs. There were a few differential brain regions on AV45-PET and FDG-PET for HC vs. low-risk MCI, but most of the brain regions showed significant difference on VBM-MRI. In the comparison of high-risk MCIs vs. ADs, brain imaging QTs of all three modalities showed a wide range of significant differences, suggesting that the conversion from high-risk MCI to AD would show a more comprehensive brain degeneration, which could be easily diagnosed. In addition, the fluid biomarkers in terms of A$\beta$42, p-tau181 and NfL were sensitive between high-risk MCIs and ADs, with the rest of the biomarkers showing no significant difference. The four cognitive domains also differed significantly on high-risk MCIs vs. ADs task, and not on HCs vs. low-risk MCIs. The $\emph{APOE}$ genotype showed significant difference on both comparison tasks, with the education level differed significantly on high-risk MCIs vs. AD task and handedness differed significantly between HC and low-risk MCI populations.
	
Finally, there are several limitations of the BigFirst at this point. First, the number of subjects used is not enough at the moment, and in the future, we will test it on larger data sets and expect to yield more important patterns. Second, our subtyping methodology is essentially determined by measuring the similarity of three different brain imaging modalities obtained from HCs and ADs. Whereas brain imaging QTs may contain incomplete information about the disease, depending on the time-point regarding the disease stage for inclusion, incorporation of more data modalities may be necessary to improve the accuracy of risk stratification. Finally, the low-risk MCI subtype does not necessarily mean an absolute stability, nor do the high-risk MCI subtype necessarily means deterioration. The proposed method, although able to distinguish individuals with different risk at the MCI stage, does not quantify the risk probability of each MCI patient, so we will quantify this uncertainty into probability in the future, which could be ore suitable for clinical practice. Finally, it is interesting to investigate the risk stratifying performance on other genetic-associated disorders besides AD. 
	
\section{Methods}\label{Methods}
\subsection{ADNI dataset}\label{ADNI}
Data used in the preparation of this article were obtained from the Alzheimer’s Disease Neuroimaging Initiative (ADNI) database (adni.loni.usc.edu). 

\subsubsection{Brain imaging scans}\label{QT_data}
There has been a number of signals early in the development of the disease that can help achieve risk identification for MCIs. Studies have shown that the amyloid deposition, brain glucose metabolism rate and brain atrophy happening consequently during the progression of AD \cite{ossenkoppele2015atrophy, chew2013fdg, hardy2003relationship}. AV45-PET can measure the amyloid deposition, FDG-PET measures the rate of glucose metabolism, and the atrophy of the brain is captured by the VBM-MRI scan. Therefore, we used three imaging modalities such as AV45-PET, FDG-PET and VBM-MRI with aim to obtain a comprehensive disease-related representations of human brain. These multi-modal imaging data were aligned to each subject’s same visit. The VBM-MRI were processed with voxel-based morphometry by SPM \cite{ashburner2000voxel}. And, every scan had been aligned to a T1-weighted template image, segmented to the gray matter (GM), the white matter (WM) and the cerebrospinal fluid (CSF) maps, normalized to the standard Montreal Neurological Institute (MNI) space as 2$\times$2$\times$2 $\text{mm}^3$ voxels, and smoothed with an 8mm FWHM kernel. Besides, the AV45-PET and FDG-PET scans were registered into the same MNI space. We further extracted region-of-interest (ROI) level measurements based on the MarsBaR automated anatomical labeling (AAL) atlas \cite{tzourio2002automated}. They were mean gray matter densities for VBM-sMRI scans, beta-amyloid depositions for AV45-PET scans and glucose utilizations for FDG-PET scans. In the experiments, the imaging measures were pre-adjusted to remove the effects of the baseline age, gender, education, and handedness by the regression weights derived from the HC subjects.

\subsubsection{Genetic variations}\label{SNP_data}
We used the whole-genome data, including 6000,000 SNPs, with aim to seek a comprehensive genetic basis underpinning ADs and MCIs. The subject-level genotyping data were generated by Human 610-Quad or OmniExpress Array (Illumina, Inc., San Diego, CA, USA) and preprocessed according to standard quality control (QC) and impact steps. All SNPs were coded by additive coding paradigm. To enable efficiency, we extracted those brain tissue related SNPs based on the GTEx consortium (version 8) because those remaining SNPs do not influencing the brain \cite{hartl2021coexpression, gtex2020gtex}. According to this database, we extracted splicing quantitative trait locus (sQTL) loci of 13 brain tissues including the amygdala, Brodmann area 24, caudate nucleus, cerebellar hemisphere, cerebellum, cortex, Brodmann area 9, hippocampus, hypothalamus, nucleus accumbens, putamen, spinal cord, and substantia nigra. We then matched these brain tissue related SNPs with the ADNI genetic variations, yielding 13 groups of SNPs of 1,517,688 loci in total. Finally, 359,997 SNPs were generated after deleting duplicates.

\subsubsection{Cognitive, clinical, fluid biomarkers}\label{Other data}
The cognitive test scores, clinical and fluid biomarkers were also downloaded from the LONI website. The fluid biomarkers included CSF biomarkers of amyloid and plasma biomarkers of p-tau 181 and NfL. The memory composite (ADNI-MEM) models based on components from the Rey Auditory Verbal Learning Test (RAVLT), Alzheimer’s Disease Assessment Scale-Cognitive Subscale (ADAS-Cog), and mini-mental status exam (MMSE). The executive function composite (ADNI-EF) models based on animal and vegetable category fluency, trail-making A and B, digit span backwards, digit symbol substitution from the revised Wechsler Adult Intelligence Scale, and circle, symbol, numbers, hands, and time items from a clock drawing task. The language composite (ADNI-LAN) models using animal and vegetable category fluency, the Boston naming total, MMSE language elements, following commands/object naming/ideational practice from ADAS-Cog, and Montreal Cognitive Assessment (MoCA) language elements, including letter fluency, naming, and repeating tasks. Further detail on these composite measures can be obtained at the ADNI website. The clinical data contained a total of 28 clinical symptoms posted on ADNI that subjects experienced during the three months prior to the baseline visit. The clinical symptoms include nausea, vomiting, diarrhea, constipation, abdominal discomfort, sweating, dizziness, low energy, drowsiness, blurred vision, headache, dry mouth, shortness of breath, coughing, palpitations, chest pain, urinary discomfort, urinary frequency, ankle swelling, musculoskeletal pain, rash, insomnia, depressed mood, crying, elevated mood, wandering, fall and other. 

\subsection{The proposed method}\label{method}
We proposed a brain imaging genetics fusion method to identify MCI risk subtypes by learning discriminative pattern of HCs and ADs. BigFirst was comprised of three modules: (1) the Risk Genetic information Extraction (RGE) module, which extracted disease-related genetic representations from extensive genetic loci; (2) the Risk Genetic Guided Pseudo-brain Generation (RG2PG) module, which leveraged the genetic representations from RGE to generate multimodal disease-specific pseudo-brain imaging template; and (3) the Cross Modality Pseudo-brain Fusion (CMPF) module, which integrated the multiple pseudo-brain guided imaging QTs using a multimodal fusion approach for disease prediction and MCI risk subtype identification.

\subsubsection{Risk Genetic information Extraction Module (RGE)}\label{RGE}
The primary function of this module was to extract disease-related genetic representations from the whole-genome. The genetic data used here are totally 13 groups, with each group corresponding to a certain brain tissue. The details were shown in Section \ref{SNP_data}. Given the inherently high-dimensional nature of human genome and the fact that not all loci are disease-relevant, it was essential to filter out genetic variations associated with the disease. Mamba, known for its efficiency in processing long one dimensional sequences, was particularly well-suited for handling the high-dimensional characteristics of genetic data \cite{gu2023mamba}. Therefore, we employed the Mamba technique as a genetic encoder to extract disease-related genetic information, i,e,
	\begin{equation}
		\mathbf{G_p}=Mamba(\mathbf{X_p}), p=1,...,13.\label{RGE_eq},
	\end{equation}
Each set of outputs is then concatenated to get the final output $\mathbf{G}=Concat([\mathbf{G_1},...,\mathbf{G_13}])$.
	
\subsubsection{Risk Genetic Guided Pseudo-brain Generation Module (RG2PG)}\label{DPG}
This module leveraged the genetic representations to identify multimodal brain imaging features, thereby enhancing gene-induced feature changes. Specifically, we employed attention masks (weights) derived from genetically generated phenotypes to select lesion brain features \cite{ma2024eye}. Given the complexity of neuroimaging data and the challenges in generating phenotypes from genetic information, we adopted an adversarial generative network (GAN), inspired by \cite{ko2022deep, ko2021engine}. By inputting genetic data and phenotypes into GAN, we generated gene-guided phenotypes, enabling the implicit learning of neuroimaging data distribution patterns. Firstly, we defined $M$ generators $G_m: \mathbb{R}^{dim(\mathbf{G})}\rightarrow\mathbb{R}^{2\times d_m}, m=1, ..., M$,
	\begin{equation}
		[\mathbf{Y_m'}, \mathbf{W_m}]=G_m(\mathbf{G}), m=1,...,M.\label{DPG_eq}
	\end{equation}
$\mathbf{Y_m'}$ was the generated phenotype and $\mathbf{W_m}$ was the attention vector of the $m$-th brain modality generated by genetic representations. Then we defined $M$ discriminators, denoted as $D_m$, to assess the reliability of the generated data. By generating multiple pseudo-brain modalities from the represented genetic data, the generator network autonomously learned the associations between brain modalities and risk genetic information. Consequently, we hypothesized that the attention vectors generated by the network have captured these associations. The loss function of the GAN adopted the least squares GAN (LSGAN) structure \cite{mao2017least}, which ensured a faster convergence and greater stability. The loss function defined as:
	\begin{equation}
		\begin{aligned}
			L_{dis}^m&=\frac{1}{2}(\mathbb{E}_{\mathbf{Y_m}}[(\mathbf{D_m}(\mathbf{Y_m})-1)^2]+\mathbb{E}_{\mathbf{Y_m'}}[(\mathbf{D_m}(\mathbf{Y_m'}))^2])\\
			L_{gen}^m&=\frac{1}{2}\mathbb{E}_{\mathbf{Y_m'}}[(\mathbf{D_m}(\mathbf{Y_m'}-1))^2].\label{GAN}
		\end{aligned}
	\end{equation}
The final output of RG2PG was $\mathbf{\hat{Y}_m}=\mathbf{Y_m}\odot\mathbf{W_m}, m=1,..,M$.
	
\subsubsection{Cross Modality Pseudo-brain Fusion (CMPF)}\label{MBIF}
The primary objective of this module was to integrate multimodal brain imaging data generated by the RG2PG module. Given the heterogeneous and multifactorial nature of AD, no single modality can comprehensively capture its complexity \cite{avelar2023decoding}. Therefore, multimodal data fusion is essential to incorporate disease-related information from diverse perspectives. We employed a supervised contrastive learning approach \cite{khosla2020supervised, yuan2023small} for multimodal data fusion to better extract discriminative features for different diagnostic groups. Specifically, for two modalities $\mathbf{\hat{Y}_m}$, $\mathbf{\hat{Y}_n}$ and their corresponding mappings $f_m, f_n: d_m\rightarrow k, d_n\rightarrow k$, we concatenate the mapped dataset as $\mathbf{Z_{m,n}}=\mathbf{Z_m}\bigcup\mathbf{Z_n}=f_m(\mathbf{\hat{Y}_m})\bigcup f_n(\mathbf{\hat{Y}_n})$. Assuming $z_{m,n}^i \in \mathbf{Z_{m,n}}$ is the anchor and the corresponding diagnostic label is $y_{m,n}^i$, the contrastive loss is
\begin{equation}
	\mathcal{L}_{cl}^{m,n}=\sum_{i\in I(i)}\frac{-1}{\left |P(i)  \right | } \sum_{p\in P(i)}\log{\frac{\exp (\mathbf{z}_{m,n}^i\cdot \mathbf{z}_{m,n}^p/\tau)}{\sum_{a\in E(i)}\exp(\mathbf{z}_{m,n}^i\cdot \mathbf{z}_{m,n}^a/\tau)} },\label{MPF}
\end{equation}
where $E(i)$ denotes the index set of all subjects except anchor, $P(i)={p\in E(i): y_{m,n}^p=y_{m,n}^i}$ means the positive set, $\tau\in\mathbb{R}^+$ is a scalar temperature parameter. If there are more than two modalities, the objective becomes:
	\begin{equation}
		\mathcal{L}_{mcl}^{m,n} = \sum_{\substack{m\in[1,...,M-1],\\
				n\in[m+1,...,M]}}\mathcal{L}_{cl}^{m,n},\label{MPF_all}
	\end{equation}
Then we predict the disease status by applying the GMM algorithm to the obtain the neuroimaging representations. The three aforementioned modules are integrated organically to form the final BigFirst method.

	
\subsection{Model settings and interpretation}\label{Interpretation}
We elaborated on the configuration details of the model parameters in Supplementary \ref{Parameters setting}. We also analyzed the parameter sensitivity and provided the results in Supplementary \ref{Parameter sensitivity analysis}. As aforementioned above, we trained the model on HC vs. AD populations. The prediction in terms of classifying HCs and ADs and related ablation results (with the REG module removed) were detailed in Supplementary \ref{Disease prediction and ablation study}. To interpret the results regarding genetic findings, we built a linear regression with risk genetic information as dependent variables and raw SNP loci as independent variables. Concurrently, we performed sparse canonical correlation analysis (SCCA) between the original brain imaging QTs and the learned multi-modal representations for brain imaging data. The interpretation results were detailed in Section \ref{high_dim_gene} and Supplementary \ref{MF interpretation}.

\section{Data avaliability}	
Data used in the preparation of this article were obtained from the Alzheimer’s Disease Neuroimaging Initiative (ADNI) database (adni.loni.usc.edu). The primary goal of ADNI has been to test whether serial magnetic resonance imaging (MRI), positron emission tomography (PET), other biological markers, and clinical and neuropsychological assessment can be combined to measure the progression of mild cognitive impairment (MCI) and early AD. For up-to-date information, see www.adni-info.org.
\backmatter
	
	%
	%
	%
	
	\bmhead{Acknowledgements}
	
	Acknowledgements are not compulsory. Where included they should be brief. Grant or contribution numbers may be acknowledged.
	
	Please refer to Journal-level guidance for any specific requirements.
	
	%
	%
	%
	
	\bigskip

\begin{appendices}
\section{Experiment supplementary}\label{secA1}
		
\subsection{Parameters setting}\label{Parameters setting}
The experiments were conducted on an NVIDIA RTX 3090 device. All experiments used the Adam optimizer with a training epoch number of 200. The learning rate was initially set to 1e-4 and decayed exponentially by $4\%$ per 1000 iterations \cite{ko2021engine, ko2022deep}. The temperature coefficient $\tau$ of contrastive learning was set to 0.07 as suggested \cite{wang2021understanding}. Batch size was fixed at 64. For the multi-modal brain imaging encoder, we used the MLP since the brain imaging data were aligned. A more complicated CNN could be also employed but required too many computational resources and was hard to interpret. In order to reduce the consumption of computational resources, we used only one Mamba block in the RGE module because SNPs data was one-dimensional. The number of channels of mamba was set to 3, corresponding to the number of encoding channels of SNP. The output dimension of mamba was set to 1. We selected the optimal parameters using a grid search method to find appropriate values for the rest of the parameters such as the dimension of the mamba block, the number of layers of the multimodal phenotype encoder, and the dimension of the multimodal phenotype representation. According to the experimental results \cite{liu2025vmamba}, the SSM state dimension of Mamba has a small impact on the performance, and thus we chose its candidate set as $\{1, 2, 4\}$ with the resource consumption taken into consideration. The candidate set of MLP structure was $\{[layer1: 128, layer2: 256], [layer1: 128, layer2: 256, layer3: 512], [layer1: 128, layer2: 256, layer3: 512], [layer1: 128, layer2: 256, layer3: 512, layer4: 1024]\}$, with the ReLU activation functions between the linear layers. The candidate set of dimensions for the multi-modal neuroimaging representations was $\{32, 64, 128\}$. Finally, the parameters were determined as follows: SSM state dimension = 2, representation dimension = 64, MLP structure = $\{[layer1: 128, layer2: 256]\}$.
		
\subsection{Model interpretation}\label{Model interpretation}
		
\subsubsection{Multi-modal neuroimaging fusion}\label{MF interpretation}
To explore the contribution of raw brain imaging QTs to the learned multi-modal representation, we conducted SCCA on the raw imaging QTs and the neuroimaging representations. The selected imaging QTs were shown in Table~\ref{SCCA_QT}. The neuroimaging QTS of AV45-PET with higher contribution were mainly come from the frontal and cingulum areas \cite{du2020associating, vanhoutte2021evaluation}. The important neuroimaging QTs of FDG-PET and VBM-MRI were mainly distributed on hippocampus and temporal, which was in line with the findings of previous studies \cite{mu2011adult, chan2001patterns, killiany1993temporal}.
		\begin{table}[h]
			\caption{Top 10 ROIs in three brain imaging modalities.}\label{SCCA_QT}%
			\begin{tabular}{@{}lll@{}}
				\toprule
				AV45-PET & FDG-PET & VBM-MRI\\
				\midrule
				Frontal-Med-Orb-Left    	&	ParaHippocampal-Left 	& Hippocampus-Right       	\\
				Cingulum-Mid-Left    		&	Hippocampus-Left  		& Hippocampus-Left     		\\
				Rectus-Left    				&	Temporal-Inf-Left  		& Temporal-Mid-Left     	\\
				Frontal-Med-Orb-Right    	&	Angular-Left  			& Temporal-Inf-Right      	\\
				Frontal-Sup-Medial-Left    	&	Cingulum-Post-Left  	& Temporal-Inf-Left     	\\
				Frontal-Mid-Orb-Right   	&	Hippocampus-Right  		& Temporal-Sup-Left    		\\
				Precuneus-Left   			&	ParaHippocampal-Right  	& Temporal-Mid-Right    	\\
				Frontal-Sup-Orb-Right   	&	Temporal-Mid-Left  		& Temporal-Sup-Right   		\\
				Frontal-Sup-Medial-Right   	&	Temporal-Inf-Right  	& Temporal-Pole-Mid-Left    \\
				Cingulum-Ant-Left   		&	Angular-Right  			& SupraMarginal-Left    	\\
				\botrule
			\end{tabular}
		\end{table}
		
		
\subsection{Parameter sensitivity analysis}\label{Parameter sensitivity analysis}
\begin{figure}[!htbp]
			\centering
			\begin{subfigure}{0.325\linewidth}
				\centering
				\includegraphics[width=0.9\linewidth]{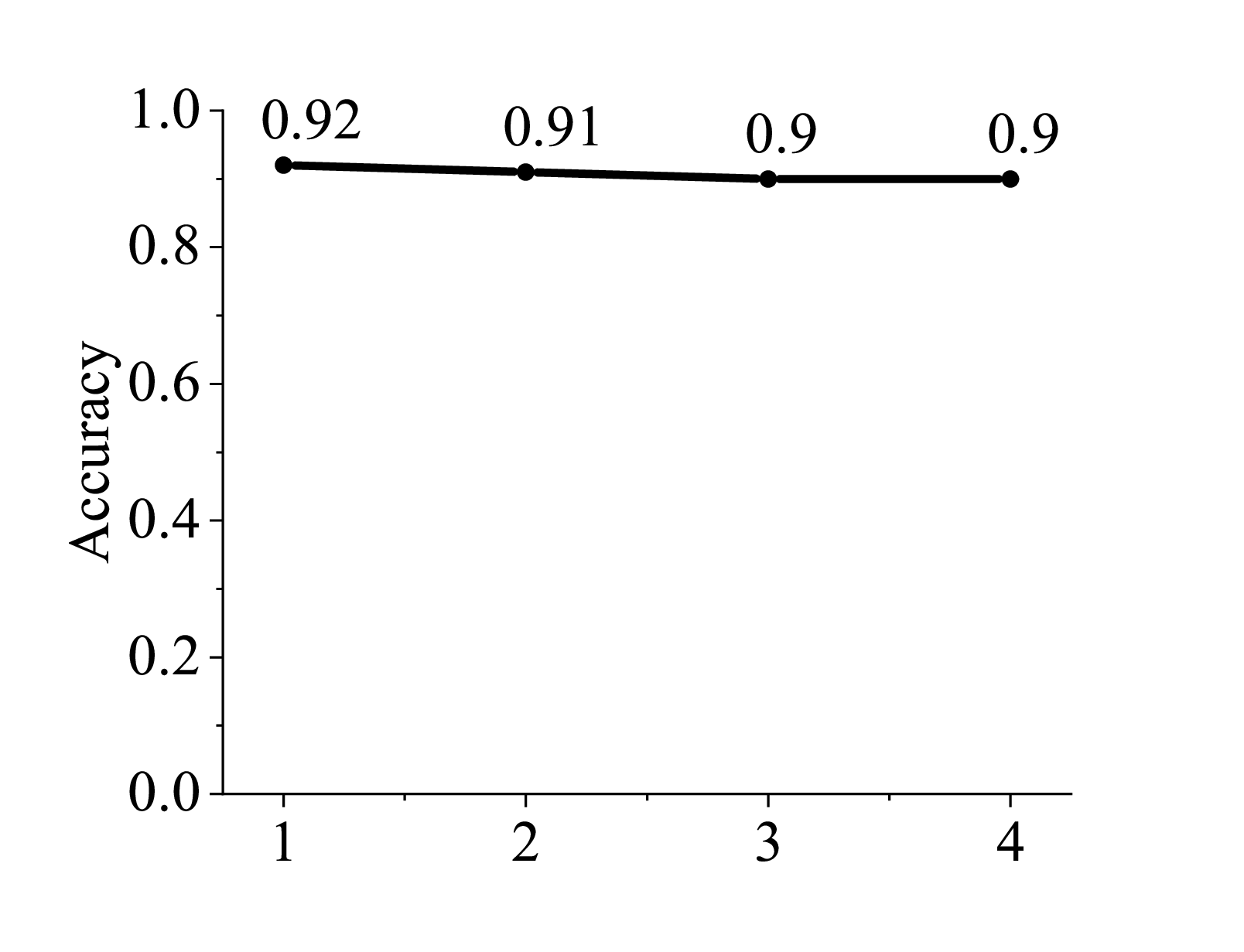}
				\caption{SSM state dimension}
				\label{SSM_sensitivity}
			\end{subfigure}
			\centering
			\begin{subfigure}{0.325\linewidth}
				\centering
				\includegraphics[width=0.9\linewidth]{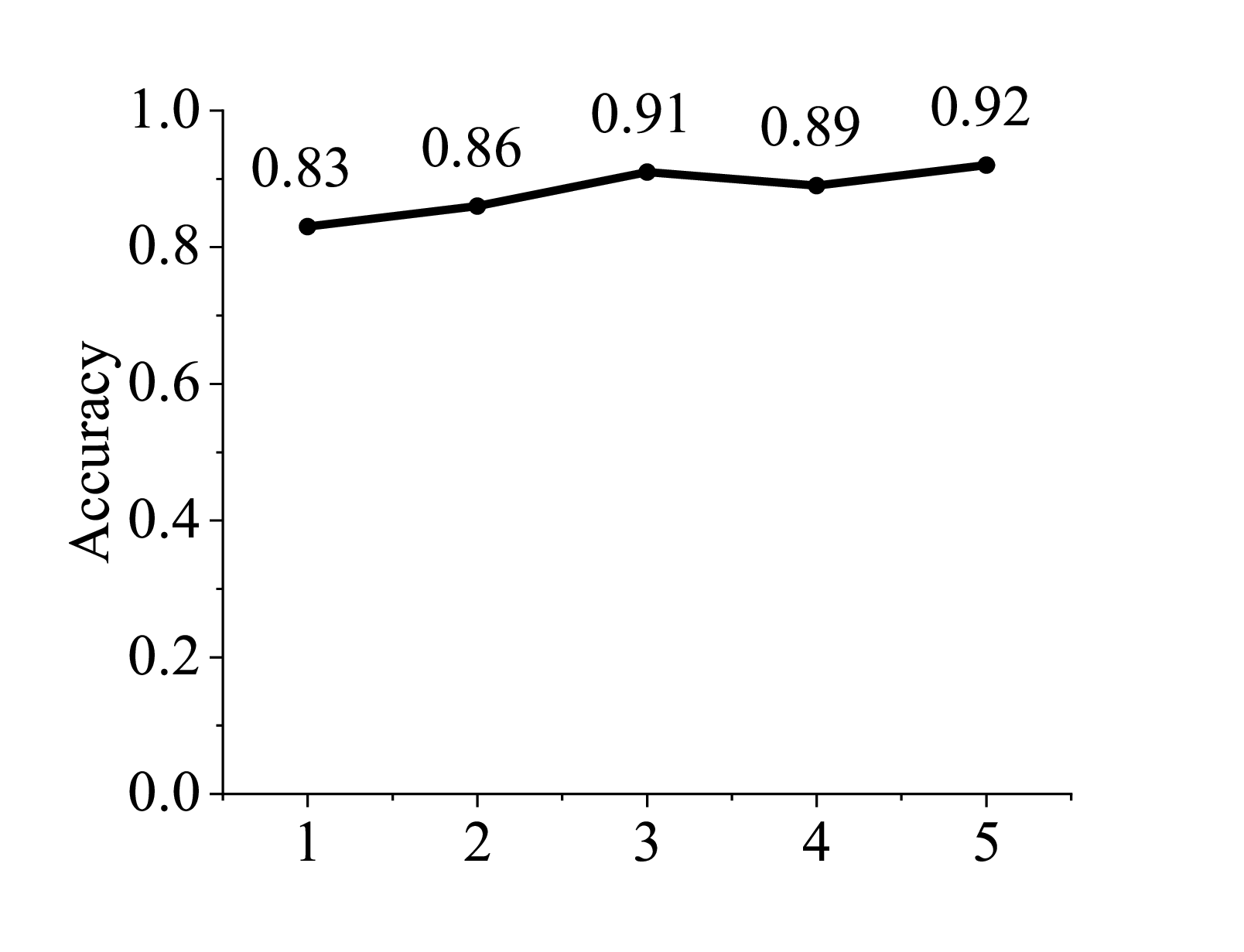}
				\caption{number of MLP layers}
				\label{layer_sensitivity}
			\end{subfigure}
			\centering
			\begin{subfigure}{0.325\linewidth}
				\centering
				\includegraphics[width=0.9\linewidth]{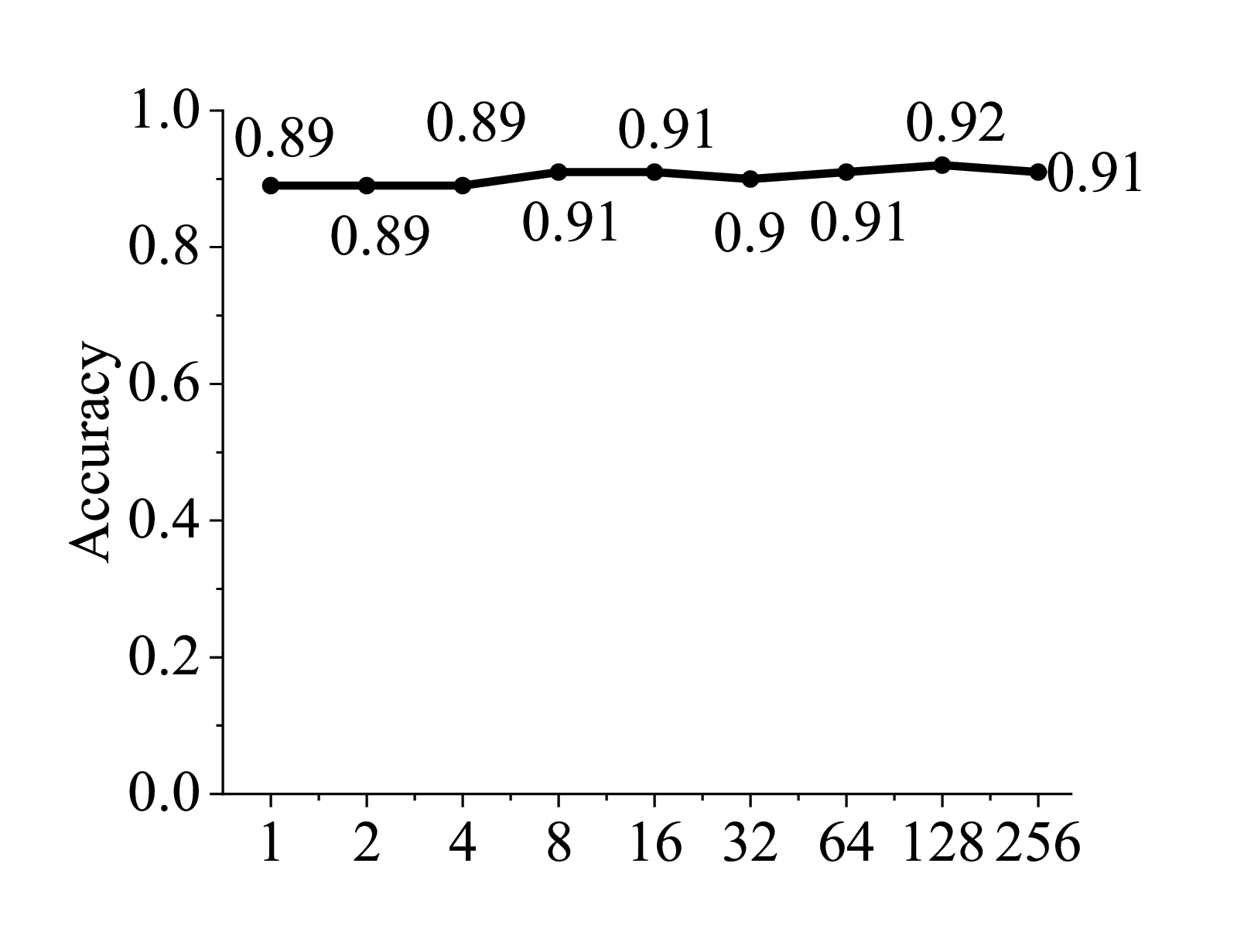}
				\caption{representation dimension}
				\label{representation_sensitivity}
			\end{subfigure}
			\caption{The curve of prediction accuracy changes as the parameters vary.}
			\label{sensitivity}
		\end{figure}

In this section, we investigated the sensitivity of three parameters, i.e., the SSM state dimension, the number of MLP layers, and the multi-modal representation dimension. Without loss of generality, we fixed two of the three parameters and varied the remaining one. The performance curve with the varying parameter was plotted in Fig.~\ref{sensitivity}. If fixed, the SSM state dimension was set to 2, the MLP layers was $\{[128, 256, 512, 1024, 2048]\}$, and the dimension of multi-modal representations was set to 128. Obviously, the results demonstrated that the prediction accuracy remains relatively stable for all three parameters, suggesting a good robustness of the BigFirst.
		
\subsection{Model performance}\label{Model performance}
\subsubsection{Disease prediction and ablation study}\label{Disease prediction and ablation study}
In this section, we presented the performance of BigFirst in classifying HCs and ADs. Training was performed using 5-fold cross-validation. The f1 score, recall, precision and accuracy were used as evaluation criteria as a classification task usually did. We further investigated the impact of the inclusion or exclusion of genetic data on model performance. Therefore, we compared the performance of the model without the RGE module to the completed model, and presented the results in the table~\ref{pred_ablation}.
		\begin{table}[h]
			\caption{Results of disease prediction and ablation study.}\label{pred_ablation}%
			\begin{tabular}{@{}lllll@{}}
				\hline
				&F1 score & Recall & Precision & Accuracy \\
				\hline
				without RGE	&	$0.88\pm 0.03$    	&	$0.87\pm 0.08$ 	& $0.89\pm 0.06$  & $0.90\pm 0.02$ \\
				with RGE	&	$0.91\pm 0.06$    	&	$0.89\pm 0.09$ 	& $0.94\pm 0.05$  & $0.93\pm 0.04$ \\
				\hline
			\end{tabular}
		\end{table}
%
		
\subsection{Covariates}\label{supp_COVS}
The distribution differences of the five covariates across four population subgroups, along with their corresponding p-values, were shown in Table~\ref{supp_COVS}.
		\begin{figure*}[ht]
			\centering
			\includegraphics[width = \linewidth]{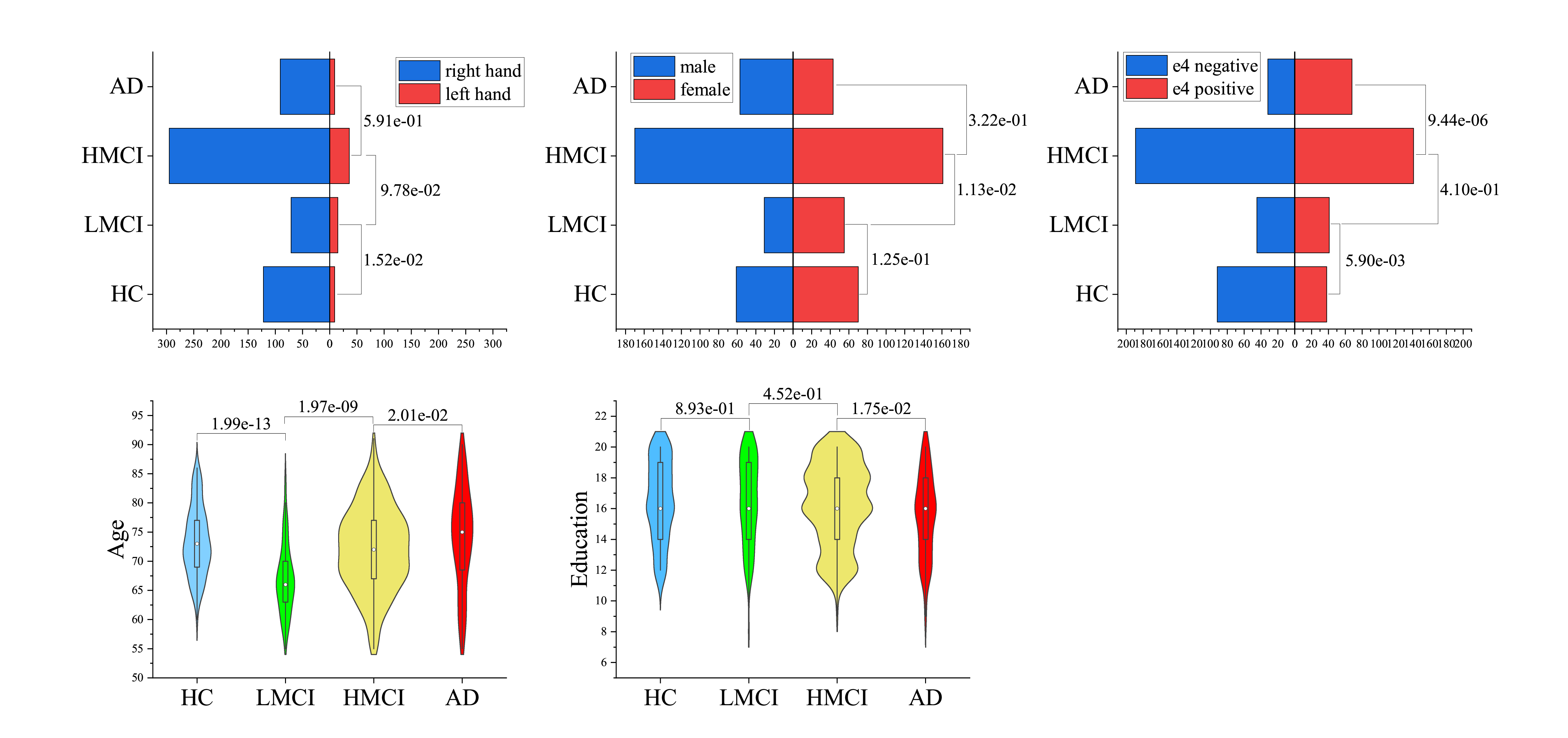}
			\caption{The distribution of different groups in five covariates with $p$-values. Top: hand, gender and $\emph{APOE}$ genotype. Bottom: Age and Education.} \label{plt_COVS}
		\end{figure*}
		
\subsection{GWAS and PheWAS}\label{supp_GWAS}
The significant SNPs identified by GWAS were shown in Table~\ref{Tab_GWAS}, and the top ten significant traits obtained by PheWAS were shown in Table~\ref{Tab_PheWAS}.
		\begin{table}[htbp!]
			\caption{Significant SNPs in GWAS.}\label{Tab_GWAS}%
			\begin{tabular}{@{}llllll@{}}
				\toprule
				Chr & SNP & Gene & A1 & OR & $p$-value\\
				\midrule
				3    &	rs75806884 & $\emph{DRD3}$     &T  & 5.520     & $5.01\times 10^{-6}$  \\
				7    &	rs6943378  & $\emph{ABCA13}$   &G  & 3.037     & $2.36\times 10^{-6}$  \\
				7    &	rs6948188  & $\emph{ABCA13}$   &A  & 3.020     & $2.85\times 10^{-6}$  \\
				9    &	rs1885426  & $\emph{HABP4}$    &T  & 2.446     & $7.60\times 10^{-6}$  \\
				9    &	rs7035310  & $\emph{CDC14B}$   &A  & 3.095     & $2.16\times 10^{-7}$  \\
				12   &	rs2108570  & $\emph{CACNA1C}$  &A  & 2.945     & $1.07\times 10^{-6}$  \\
				12   &	rs2283296  & $\emph{CACNA1C}$  &A  & 2.639     & $2.52\times 10^{-6}$  \\
				12   &	rs2238060  & $\emph{CACNA1C}$  &C  & 2.639     & $2.52\times 10^{-6}$  \\
				12   &	rs2239041  & $\emph{CACNA1C}$  &A  & 2.843     & $8.30\times 10^{-7}$  \\
				12   &	rs2239042  & $\emph{CACNA1C}$  &G  & 2.814     & $1.10\times 10^{-6}$  \\
				12   &	rs4765919  & $\emph{CACNA1C}$  &T  & 2.683     & $1.70\times 10^{-6}$  \\
				12   &	rs2239045  & $\emph{CACNA1C}$  &T  & 2.683     & $1.70\times 10^{-6}$  \\
				12   &	rs2239046  & $\emph{CACNA1C}$  &G  & 2.619     & $2.44\times 10^{-6}$  \\
				12   &	rs3819531  & $\emph{CACNA1C}$  &C  & 2.518     & $7.21\times 10^{-6}$  \\
				12   &	rs3819536  & $\emph{CACNA1C}$  &G  & 2.518     & $7.21\times 10^{-6}$  \\
				12   &	rs1544503  & $\emph{CACNA1C}$  &A  & 2.675     & $3.88\times 10^{-6}$  \\
				12   &	rs2238065  & $\emph{CACNA1C}$  &A  & 2.675     & $3.88\times 10^{-6}$  \\
				12   &	rs2238066  & $\emph{CACNA1C}$  &G  & 2.518     & $7.21\times 10^{-6}$  \\
				15   &	rs12324698 & $\emph{NARG2}$    &C  & 2.694     & $3.64\times 10^{-6}$  \\
				18   &	rs80051604 & $\emph{C18orf34}$ &T  & 3.597     & $1.62\times 10^{-7}$  \\
				\botrule
			\end{tabular}
			\footnotetext{Chr: chromosome. SNP: rsids of the SNPs. A1: effect allele. OR: odds ratio values. Gene: nearest gene. $p$-value: GWAS $p$-value.}
		\end{table}
		
		\begin{table}[htbp!]
			\caption{Top ten significant traits of $\emph{CACNA1C}$ in PheWAS.}\label{Tab_PheWAS}%
			\begin{tabular}{@{}lll@{}}
				\toprule
				Domain & Trait & $p$-value\\
				\midrule
				Psychiatric    	&	Schizophrenia/Bipolar disorder & $1.66\times 10^{-23}$  \\
				Cardiovascular 	&	Pulse rate (automated reading)  & $1.45\times 10^{-22}$  \\
				Psychiatric    	&	Schizophrenia  & $3.88\times 10^{-22}$  \\
				Psychiatric    	&	Schizophrenia  & $9.53\times 10^{-21}$  \\
				Psychiatric    	&	Schizophrenia vs Bipolar disorder  & $9.53\times 10^{-21}$  \\
				Cardiovascular 	&	Resting heart rate  & $1.97\times 10^{-20}$  \\
				Metabolic   	&	Body Mass Index  & $7.81\times 10^{-16}$  \\
				Psychiatric   	&	Schizophrenia/Bipolar disorder  & $1.04\times 10^{-14}$  \\
				Psychiatric   	&	Schizophrenia  & $1.83\times 10^{-14}$  \\
				Psychiatric   	&	Schizophrenia  & $9.26\times 10^{-14}$  \\
				\botrule
			\end{tabular}
		\end{table}
		
		\begin{figure}[H]
			\centering
			\includegraphics[width = 0.6\linewidth]{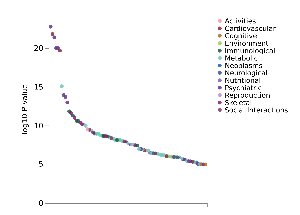}
			\caption{PheWAS plot of gene $\emph{CACNA1C}$.} \label{Fig_Phewas_CACNA1C}
		\end{figure}
		
		
		
		
	\end{appendices}
	
	
	\bibliography{sn-bibliography}

\end{document}